\documentclass[reprint,amsmath,amssymb,aps,floatfix,pre,superscriptaddress]{revtex4-1}

\usepackage{savesym}
\usepackage{graphicx}
\usepackage{dcolumn}
\usepackage{bm}
\usepackage{epstopdf}
\usepackage{pifont}
\usepackage{xcolor}
\usepackage{blindtext}

\savesymbol{corresponds}
\usepackage{mathabx}

\begin{document}

\preprint{}

\title{Anomalous behavior of plasma response functions at strong coupling }

\author{P\'eter Magyar}
\affiliation{Institute for Solid State Physics and Optics, Wigner Research Centre for Physics, P.O.B. 49, H-1525 Budapest, Hungary}

\author{Gabor J. Kalman}
\affiliation{\ Department of Physics, Boston College, Chestnut Hill, MA 20467, USA}

\author{P\'eter Hartmann}
\affiliation{Institute for Solid State Physics and Optics, Wigner Research Centre for Physics, P.O.B. 49, H-1525 Budapest, Hungary}

\author{Zolt\'an Donk\'o}
\affiliation{Institute for Solid State Physics and Optics, Wigner Research Centre for Physics, P.O.B. 49, H-1525 Budapest, Hungary}
\date\today
\begin{abstract}
Using data from equilibrium Molecular Dynamics computer simulations we have built up a catalogue  of response functions for the Coulomb  one-component plasma (OCP) over a wide range of $\Gamma$ coupling values, including the strongly coupled $\Gamma > 1$ liquid regime. We focus on the domain of negative compressibility ($\Gamma>3$), where causality and the concomitant Kramers-Kronig relations are violated. We give a description of the details of the violation, both in the frequency and time domains. We show that the viscoelastic pole of the density response function morphs into an imaginary pole in the upper $\omega$ half-plane that is responsible for the anomalous behavior of the response in this coupling range. By examining the plasmon dispersion relation through the dielectric response function, rather than via the  peaks of the dynamical structure function, we obtain a more reliable representation for the dispersion. We demonstrate  that there is an intimate link between the formation of the roton minimum in the dispersion and the negative compressibility of the system. The feasibility of the extension of our analysis to systems with a short range interaction is explored.
\end{abstract}

\pacs{52.27.Lw, 81.40.Lm, 62.20.Hg}

\maketitle

Linear response functions play a pivotal role in the theory of classical and quantum  many particle systems. Expressed in the language of frequency $(\omega)$ and wave vector $(\textbf{k})$ dependence, they reveal a wealth of information
both about  the static and the dynamic properties of  the system, most importantly  about its collective excitation spectrum.
The response functions  are linked to the equilibrium properties of the system through the profound Fluctuation Dissipation Theorem (FDT) \cite{PN,Ichimaru}. At the same time, the functional dependence on the $(\textbf{k},\omega)$ variables is limited by a number of so-called sum rules \cite{PN,ADNS}, which then can serve either as a guidance or a control  on adopting approximation schemes for actual calculations. An additional constraint   established as a consequence of the postulated causal behavior of the response function (derived from identifying the response of the system as the "effect" due to the perturbation by an external agent as the  "cause")
 is known as Kramers-Kronig (KK) relations:
a detailed discussion on this feature will be a central theme of the present work.\par

Immense efforts have been spent since the 1950-s on the calculation of plasma response functions, both for classical plasmas \cite{Ichimaru} and  for the degenerate electron gas in condensed matter \cite{PN}. (For an up-to-date summary see \cite{Vignale}). Of central interest is the question how the behavior of the response  functions  is affected by interparticle correlations, i. e. the strength of the coupling in the plasma.
Coupling is determined by the ratio of the potential and the kinetic energies of the particles and is routinely characterized by the parameters $\Gamma=Z^{2}e^{2}\beta/a$, $\beta=1/k_{\rm B}T$ or $r_{S}=a/a_{\rm Bohr}$ for classical or for degenerate quantum systems, respectively  ($a=(3/4\pi n_{0})^{1/3}$ is the Wigner-Seitz radius, $n_0$ is the density of the homogeneous system and $Z$ is the charge number). Most of the existing calculations have addressed the weak coupling regime where correlations are negligibly small and where the Vlasov or Random Phase Approximation (RPA) \cite{PN,Ichimaru}, based precisely on the complete disregard of correlations, has been used successfully. Systematic perturbation calculations for classical plasma's with finite but still weak correlations have been pursued \cite{Dawson,Coste,Dubois}, but led only to results of great complexity and of limited physical insight.  For stronger coupling,  no reliable calculations, except through some rather drastic approximations, \cite{STLS,QLCA} and justifiable only within restricted domains of the  $(\textbf{k},\omega)$. space, are available.

In this paper, we approach the problem of exploring the algebraic structure and physical contents of response functions for classical plasmas at intermediate and strong coupling values from a different angle. Recent progress in the computer simulation of the equilibrium dynamics of plasmas
has resulted in the availability of high quality data for the various  equilibrium dynamical fluctuation spectra of the system in a wide range of coupling values \cite{Ihor}. By exploiting the FDT, these data can be converted, as explained below, into the complete description of the plasma response functions of interest. A full account  of our investigations along this line will be reported elsewhere. Here we use the new information gathered to study a more specific old problem that relates  to the behavior  at higher coupling values: it is the long-standing issue of the apparent violation of causality and the concomitant violation of the KK relations for a certain class of response functions, which is the focus of the present work.

The plan of the paper is as follows. In Section I we review the historical background of the problem  and describe our approach to its rigorous treatment.  In Section II we present our results  on the detailed description of the violation, and give a quantitative discussion of its strong dependence on the coupling strength.
We also  identify the minimum coupling value  for the violation to occur and delineate the region in the $(\textbf{k},\omega)$-space where the violation actually takes place.   We point out that one can identify an anomalous part of the response function
generated by a complex pole in the $\omega$-plane associated with the violation and we review the effect of the anomalous part on the formulation of the customary sum rules.  Section III is devoted to the description of the  somewhat unexpected influence of the anomalous part on the plasmon dispersion, in particular on the formation of the so-called roton minimum \cite{Kyrkos}.  In Section IV we present a simple two-pole model in order to guide in a more intuitive fashion through  the   preceding discussions and to elucidate their physical contents. Section V serves a somewhat similar purpose, complementing the discussion from a different point of view. There we explore the behavior of the violating response in the time domain, to see more precisely what  the violation of causality implies for the observable physical variables. Finally, in Section VI we raise the question as to what extent the findings of this paper are relevant to systems governed by short range forces, rather than the long-range Coulomb interaction.

\section {PURPOSE AND METHOD}

The validity of the KK relationships is based upon the response function being a causal function in the time domain (i.e. the response vanishing for times preceding the time of the perturbation) and, correspondingly, a {\it plus-function} in the frequency domain \cite{Balescu} (i.e. an analytic function in the upper frequency half plane).

Not all response functions are causal, though. This has been clearly established in the the 1960-s through the pioneering works of Martin \cite{Martin} and Kirzhnits and collaborators \cite{Kirzhnits67,Kirzhnits81}. The issue hinges upon the precise definition of the "perturbing field". Contemplating now a many particle system consisting of charged particles (a plasma) perturbed by an external electric potential $\widehat{\Phi}$, one has to differentiate between this latter and the total average electric potential $\overline{\Phi}$, consisting of $\widehat{\Phi}$ and the average polarization field generated by the system itself $\widecheck{\Phi}$,
\begin{equation}
\overline{\Phi}=\widehat{\Phi}+\widecheck{\Phi}. \nonumber
\end{equation}
Even though it is the $\overline{\Phi}$ field that the particles experience, it is only $\widehat{\Phi}$ that is under the experimentalist's control with the ability to impose an arbitrary time dependence on it. Therefore response functions that relate to the $\widehat{\Phi}$ as the perturbing field are {\it bona fide} causal functions, while those that relate to $\overline{\Phi}$ do not necessarily exhibit a causal behavior (although under certain conditions they may do so).

For plasmas there exists a family of closely related response functions \cite{PN,GKS}, which can be classified according to the above criteria to belong to either of the two groups. The inverse dielectric function $\eta(\textbf{k},\omega)$, the (external) density response function $\chi(\textbf{k},\omega)$, the (external) conductivity $\sigma(\textbf{k},\omega)$ belong to the first group, while the dielectric function $\varepsilon(\textbf{k},\omega)$, the proper density response function $\overline{\chi}(\textbf{k},\omega)$, the conductivity $\overline{\sigma}(\textbf{k},\omega)$, the polarization function $\pi(\textbf{k},\omega)$ fall in the second category. It is this violating second group of interrelated response functions, whose features carry the most direct imprint of the fundamental dynamics of the system which are of interest in this paper.

We consider a one-component plasma (OCP), consisting of classical charged particles of the same kind, embedded in a neutralizing background. The system is fully characterized by the plasma frequency, $\omega_{\rm p}=(4\pi Z^{2}e^{2}n_0/m)^{1/2}$ and the coupling constant $\Gamma$.
The (external) density response function of a plasma ${\chi}(\textbf{k},\omega)$ is defined by the relationship
\begin{equation}
n(\textbf{k},\omega)={\chi}(\textbf{k},\omega)\widehat{\Phi}(\textbf{k},\omega).
\end{equation}
In a similar fashion the proper density response function $\overline{\chi}(\textbf{k},\omega)$ is provided by
\begin{equation}
n(\textbf{k},\omega)=\overline{\chi}(\textbf{k},\omega)\overline{\Phi}(\textbf{k},\omega).
\end{equation}

At this  point we  also recall the connection with the dielectric function
\begin{align}
\varepsilon(\textbf{k},\omega)&=1-\varphi(\textbf{k})\overline{\chi}(\textbf{k},\omega),\\
\eta(\textbf{k},\omega)&=1+\varphi(\textbf{k})\chi(\textbf{k},\omega).
\end{align}
In these relationships, $ n(\textbf{k},\omega)$ is the first order perturbed density and $\varphi(\textbf{k})=4\pi  Z^{2}e^{2}/k^{2}$ is the Fourier transform of the Coulomb potential.
The response functions are complex quantities, e.g.,  ${\chi}(\textbf{k},\omega)={\chi}'(\textbf{k},\omega)+i{\chi}''(\textbf{k},\omega)$, etc. For a causal function, such as ${\chi}(\textbf{k},\omega)$, the KK relations connect the imaginary and real parts through
\begin{align}
\chi'(\textbf{k},\omega)&=\frac{1}{\pi}\mathcal{P}\int_{-\infty}^{\infty}
\frac{\chi''(\textbf{k},\nu)}{\nu-\omega}d\nu \nonumber \\
&=\frac{2}{\pi}\mathcal{P}\int_{0}^{\infty}\nu
\frac{\chi''(\textbf{k},\nu)}
{\nu^2-\omega^2}d\nu\\
\chi''(\textbf{k},\omega)&=-\frac{1}{\pi}\mathcal{P}\int_{-\infty}^{\infty}
\frac{\chi'(\textbf{k},\nu)}{\nu-\omega}d\nu \nonumber \\
&=-\frac{2}{\pi}\omega\mathcal{P}\int_{0}^{\infty}
\frac{\chi'(\textbf{k},\nu)}
{\nu^2-\omega^2}d\nu
\end{align}
where $\mathcal{P}$ denotes the Cauchy principal value.
In contrast, as pointed out above, the proper density response function $\overline{\chi}(\textbf{k},\omega)$ may violate these relationships.

As discussed in the Introduction, the objective of this work is to provide an exact description of the violation, relying on equilibrium data provided by computer simulations. High quality data for the dynamical structure function
\begin{equation}
S(\textbf{k},\omega)=\frac{1}{2\pi N}\int_{-\infty}^{\infty}
\langle n_{\textbf{k}}(t)n_{-\textbf{k}}(0)\rangle^{(0)}{\rm e}^{i\omega t}dt
\label{Skw}
\end{equation}
and independently for the equilibrium static structure function
\begin{equation}
S(\textbf{k})=\frac{1}{N}
\langle n_{\textbf{k}}(0)n_{-\textbf{k}}(0)\rangle^{(0)}
\label{Sk}
\end{equation}
where $n_{\textbf{k}}(t)=\sum_{j=1}^{N}{\rm e}^{-i\textbf{k} \cdot \textbf{r}_{j}(t)}$ is the microscopic density in Fourier space have  become  by now available through Molecular Dynamics (MD) computer simulations for a wide range of parameter values, e.g., \cite{Ihor,Desbiens,Mithen,Arkhipov} and are  extended here to cover the entire domain of interest in parameter space. 

In the present simulations, we  trace a single species of $N$=10000 charged particles within a cubic box with periodic boundary conditions. To fully account for the long-range Coulomb interparticle potential we use the Particle-Particle Particle-Mesh (P$^{3}$M) \cite{Eastwood} in the calculation of the forces acting on the particles. The integration of the equations of motion of the particles is performed using the velocity-Verlet scheme. To compute the dynamical structure function we use the expression \cite{Hansen-paper}
\begin{equation}
S(\textbf{k},\omega)=\frac{1}{2\pi N}\lim_{T\rightarrow\infty}\frac{1}{T}|n_{\textbf{k}}^{\rm{FL}}(\omega)|^{2},
\label{}
\end{equation}
which is equivalent to the definition (\ref{Skw}), but it involves time average instead of ensemble average and where
\begin{equation}
n_{\textbf{k}}^{\rm{FL}}(\omega)=\lim_{T\rightarrow\infty}\int_{0}^{T}n_{\textbf{k}}(t)e^{i\omega t}dt
\label{}
\end{equation}
is the Fourier-Laplace transform of the microscopic density. This transform is carried out numerically based on the simulation measurement of the value of the fluctuating $n_{\textbf{k}}(t)$ during a sequence consisting of $N_{t}=75600$ time-steps. To improve the signal to noise ratio of the $S(\textbf{k},\omega)$ data we compute the average of $n_{\textbf{k}}^{\rm{FL}}(\omega)$ resulting from more sequences. The finite $T=N_{t}\Delta t$ time-length of the sequences provides $\Delta\omega=2\pi/T=\omega_{\rm{p}}/400$ frequency resolution ($\Delta t$ is the time-step of the simulation). Beside the dynamical structure function, we also measure its static counterpart $S(\textbf{k})$, using again time-average along the phase space trajectories of the particles. This quantity can also be computed as
\begin{equation}
S(\textbf{k})=\int_{-\infty}^{\infty}S(\textbf{k},\omega)d\omega.
\label{}
\end{equation}
The comparison of the value of this integral with the $S(\textbf{k})$ obtained directly from the simulation shows a perfect agreement, which verifies the consistency of our calculations. Because of the isotropy of the system the structure functions depend only on the absolute value of their wave-number vector argument, which is the multiples of $k_{\rm{min}}=2\pi/l$, or in normalized units $k_{\rm{min}}a=0.181$, where $l$ is the length of the edge of the cubic simulation box. Our studies cover a set of $\Gamma$ values in the strongly coupled liquid phase (1$<\Gamma <$160).

In the forthcoming calculations we need the application of the KK relations. To avoid the problem of the numerical implementation of the principal value integral appearing in the relations we use a method based on a double Fourier transformation \cite{FKK}. To describe this method let us consider a causal response function $\Psi(\textbf{k},t)$. The causality implies that $\Psi(\textbf{k},t\leq 0)=0$. The frequency-representation of the response function is given via Fourier-transform:
\begin{equation}
\Psi(\textbf{k},\omega)=\int_{-\infty}^{\infty}\Psi(\textbf{k},t){\rm e}^{i\omega t}dt.
\label{Fourier}
\end{equation}
To obtain a relation between the real and imaginary parts of $\Psi(\textbf{k},\omega)$ we split $\Psi(\textbf{k},t)$ into even and odd parts:
\begin{align}
&\Psi(\textbf{k},t) \nonumber \\
&=\frac{\Psi(\textbf{k},t)+\Psi(\textbf{k},-t)}{2}+\frac{\Psi(\textbf{k},t)-\Psi(\textbf{k},-t)}{2} \nonumber \\
&=s(\textbf{k},t)+q(\textbf{k},t),
\end{align}
where $s(\textbf{k},-t)=s(\textbf{k},t)$ and $q(\textbf{k},-t)=-q(\textbf{k},t)$. Using this partition in the Fourier-transform (\ref{Fourier}) we get
\begin{eqnarray}
\Psi'(\textbf{k},\omega)&=&2\int_{0}^{\infty}s(\textbf{k},t)\cos(\omega t)dt, \\
\Psi''(\textbf{k},\omega)&=&2\int_{0}^{\infty}q(\textbf{k},t)\sin(\omega t)dt,
\end{eqnarray}
for the real and imaginary parts of $\Psi$, respectively.

On the other hand, we know that for $t\geq 0$ the equality $s(\textbf{k},t)=q(\textbf{k},t)$ stands because of the causality of $\Psi(\textbf{k},t)$, therefore $\Psi'(\textbf{k},\omega)$ and $\Psi''(\textbf{k},\omega)$ are the cosine- and sine-transforms of the same function, respectively. In this representation  the real (imaginary) part can be computed as the cosine-transform (sine-transform) of the inverse sine-transform (cosine-transform) of the imaginary (real) part. Explicitly this takes the form
\begin{equation}
\Psi'(\textbf{k},\omega)=\frac{1}{\pi}\int_{0}^{\infty}\left[\int_{-\infty}^{\infty}\Psi''(\textbf{k},\nu)
\sin(\nu t)d\nu\right]\cos(\omega t)dt
\label{FKK1}
\end{equation}
and its counterpart is
\begin{equation}
\Psi''(\textbf{k},\omega)=\frac{1}{\pi}\int_{0}^{\infty}\left[\int_{-\infty}^{\infty}\Psi'(\textbf{k},\nu)
\cos(\nu t)d\nu\right]\sin(\omega t)dt.
\label{FKK2}
\end{equation}
Taking into account that
\begin{eqnarray}
\int_{0}^{\infty}\sin(\nu t)\cos(\omega t)dt&=&\lim_{\xi\rightarrow 0}\int_{0}^{\infty}
{\rm e}^{-\xi t}\sin(\nu t)\cos(\omega t)dt \nonumber \\
&=&\mathcal{P}\frac{\nu}{\nu^{2}-\omega^{2}},
\label{}
\end{eqnarray}
we can see that the Eqs. (\ref{FKK1}) and (\ref{FKK2}) are equivalent to the KK relations, but they have the advantage that they help to avoid the actual calculation of the unwieldy Principal Part integrals. 

It is now possible to obtain  the  $\overline{\chi}(\textbf{k},\omega)$ from the MD generated $S(\textbf{k},\omega)$ through a few simple steps. First,
by invoking the Fluctuation-Dissipation Theorem (FDT)
\begin{equation}
S(\textbf{k},\omega)=-\frac{1}{\pi\beta n_0\omega}\chi''(\textbf{k},\omega),
\label{}
\end{equation}
we acquire $\chi''(\textbf{k},\omega)$. Then, the application of the KK relation (recall that  $\chi(\textbf{k},\omega)$ is a causal function) provides the full $\chi(\textbf{k},\omega)$. Finally, the well-known relationship
\begin{equation}
\overline{\chi}(\textbf{k},\omega)=\frac{\chi(\textbf{k},\omega)}{1+\varphi(\textbf{k})\chi(\textbf{k},\omega)}
\label{chi-tot}
\end{equation}
leads to the desired result. For the static limit we use the notation $\overline{\chi}(\textbf{k},\omega=0)=\overline{\chi}(\textbf{k})$, with a similar convention for other response functions. Due to the reality condition they obey, these static responses are real functions.  Their value can be independently verified by the data for the static structure function $S(\textbf{k})$, via the static version of the FDT:
\begin{equation}
S(\textbf{k})=-\frac{1}{\beta n_0}\chi(\textbf{k}).
\label{static-FDT}
\end{equation}

\section {EXPLICIT FORMULATION OF THE VIOLATION}
In the following we study the most fundamental quantity, the proper density response $\overline{\chi}(\textbf{k},\omega)$. The implications for the other related acausal response functions of the relationships we are about to derive can be easily established. \par

The long wavelength limit of the response $\overline{\chi}(\textbf{k})$ is governed by the compressibility sum rule
\begin{equation}
\overline{\chi}_{0}\equiv\overline{\chi}(k\rightarrow 0)=-\frac{\beta n_0}{L},
\qquad L=\frac{{\partial P} / {\partial n}}{{\partial P_{0}}/{\partial n}}.
\label{compsumrule}
\end{equation}
where $n$ is the density, $P$ is the pressure, $P_0$ is the ideal gas pressure and $L=1+ L_{\rm corr}$ is the normalized inverse compressibility
(stiffness) with $L_{\rm corr}<0$ being the correlational contribution to $L$.

\begin{figure}[htb]
\includegraphics[width=1.0\columnwidth]{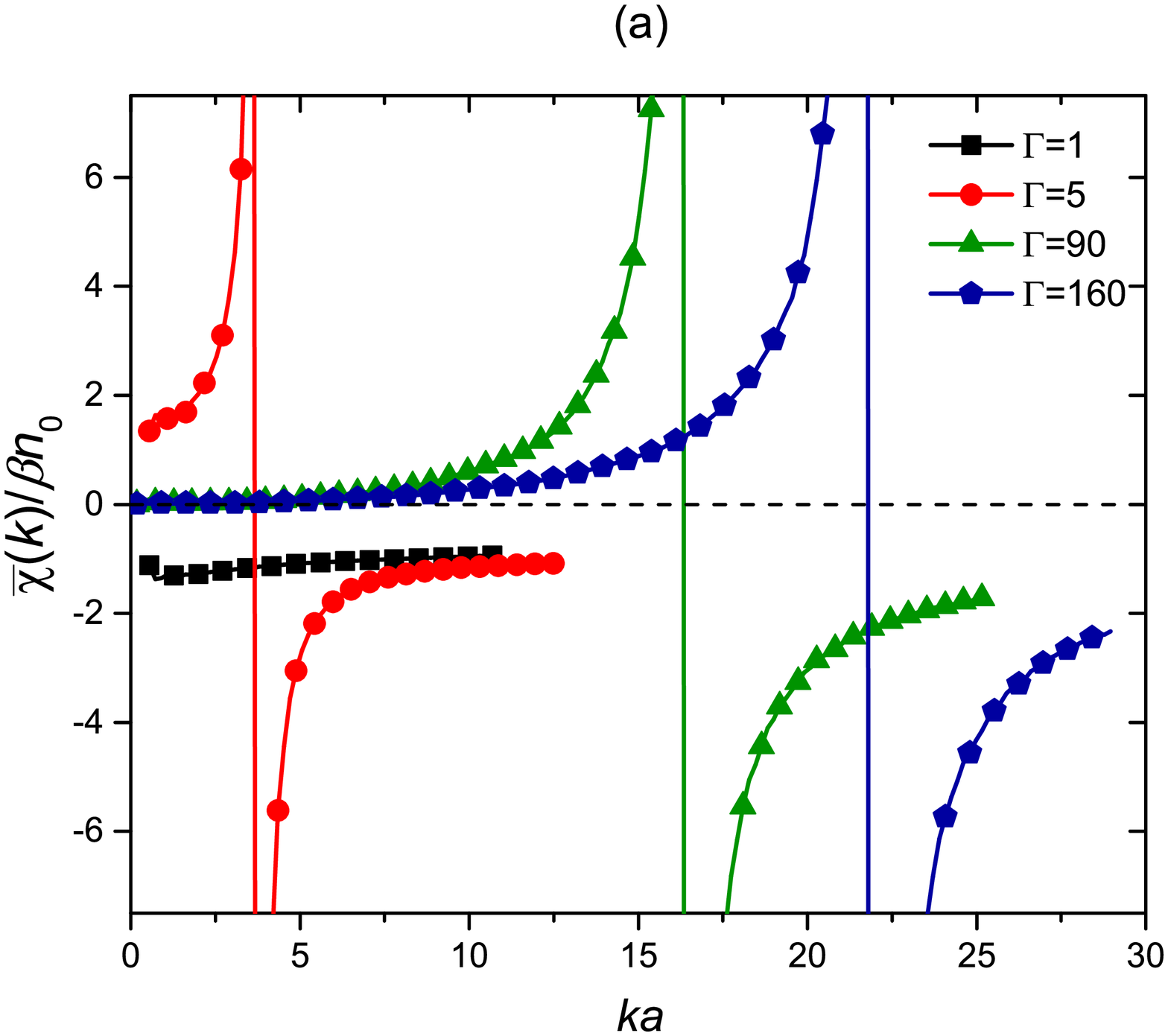}
\includegraphics[width=1.0\columnwidth]{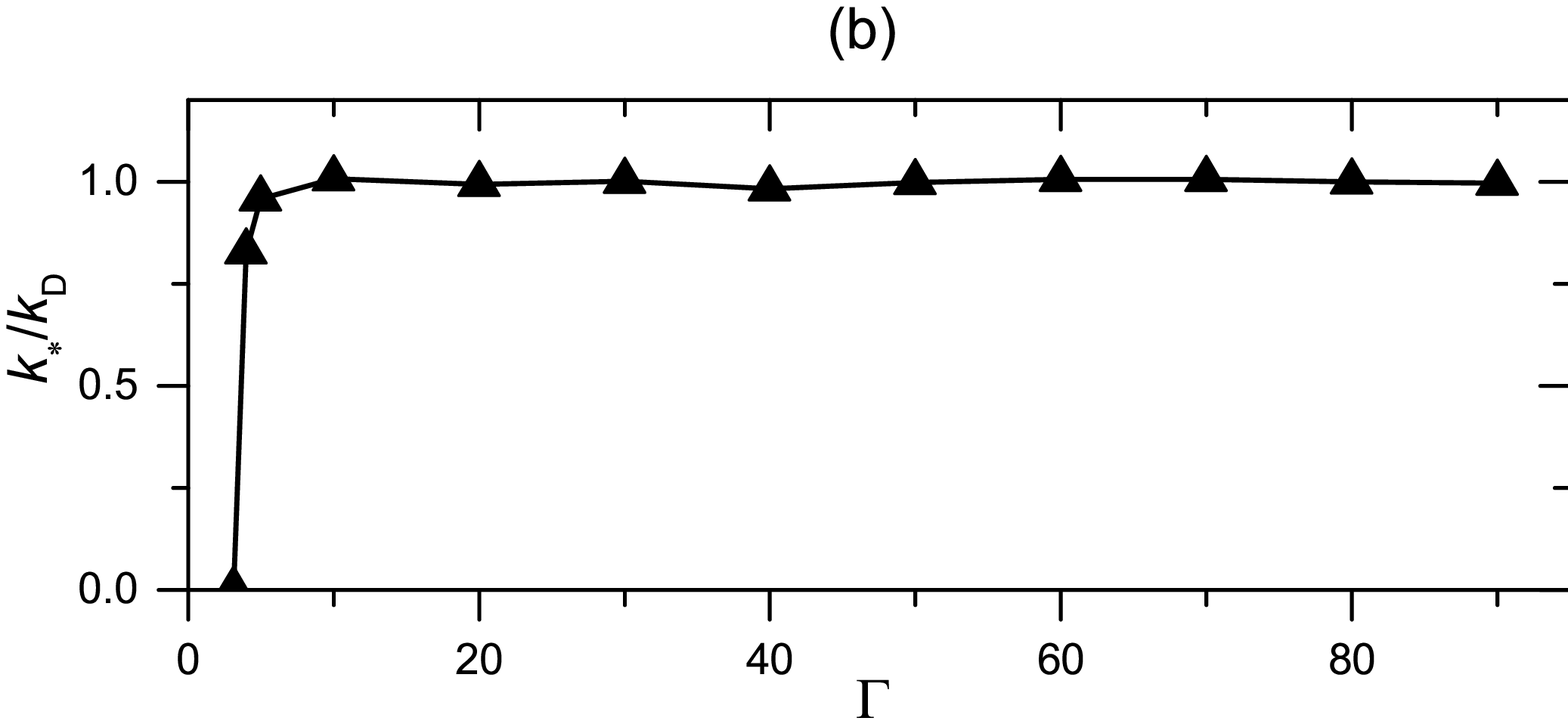}
\caption{(a) The static proper density response function $\overline{\chi}(\textbf{k})$ for $\Gamma$ values indicated. (b) Dependence of the critical wave-number $k_\ast$ on the coupling parameter $\Gamma$. Its value is normalized to the Debye wave-number $k_{\rm D}$, $k_{\rm D}a=\sqrt{3\Gamma}$.}
\label{stat}
\end{figure}

At weak coupling, even though  $L<1$, it remains positive, similarly to its behavior in the ideal gas limit. The crucial feature now  is, however, that with increasing coupling strength the compressibility changes from positive to negative around $\Gamma=\Gamma_{\ast}\simeq 3$ \cite{EOS,Khrapak-2}. (Similarly, the compressibility of a degenerate electron liquid becomes negative around $r_{s\ast}= 5.2$ \cite{Schakel,Dolgov-Maksimov,Dolgov-supercond}). Accordingly, $\overline{\chi}_{0}$ changes from negative to positive at the same point. The resulting $\overline{\chi}_{0}>0$ is incompatible with the relevant KK relation, which would require that $\overline{\chi}(\textbf{k})$ be determined by the integral
\begin{equation}
\frac{2}{\pi}\mathcal{P}\int_{0}^{\infty}\frac{\overline{\chi}''(\textbf{k},\nu)}{\nu}d\nu.
\label{integral}
\end{equation}
A positive value of this integral, however, is impossible since $\overline{\chi}''(\textbf{k},\omega)<0$ is required to ensure that $\varepsilon''(\textbf{k},\omega)$, a quantity governing dissipation, is positive. Thus for $\Gamma>\Gamma_{\ast}$,  $\overline{\chi}(\textbf{k},\omega)$ violates the KK relations and is not a $plus$-function \cite{Martin,Kirzhnits67,Mezincescu}. This violation extends beyond $k=0$ over a finite range of $0<k<k_{\ast}$ values, $k_{\ast}$ being the point where $\overline{\chi}(\textbf{k})$ reverts to its normal negative value. The details of the way this happens are depicted in Fig.~\ref{stat}(a), constructed from the MD data for $\overline{\chi}(\textbf{k},\omega)$, as well independently from $S(\textbf{k})$.
Since $\overline{\chi}(k_\ast)\to\infty$, its value can be determined from
\begin{equation}
\beta n_0 \varphi(k_{\ast})S(k_{\ast})=1.
\end{equation}
The resulting $k_{\ast}(\Gamma)$ dependence is also displayed in Fig.~\ref{stat}(b). In this figure, $k_\ast$ is normalized by the Debye wave number $k_{\rm D} = \sqrt{3 \Gamma} / a$ \cite{Ichimaru}.

\begin{figure}[htb]
\includegraphics[width=1.0\columnwidth]{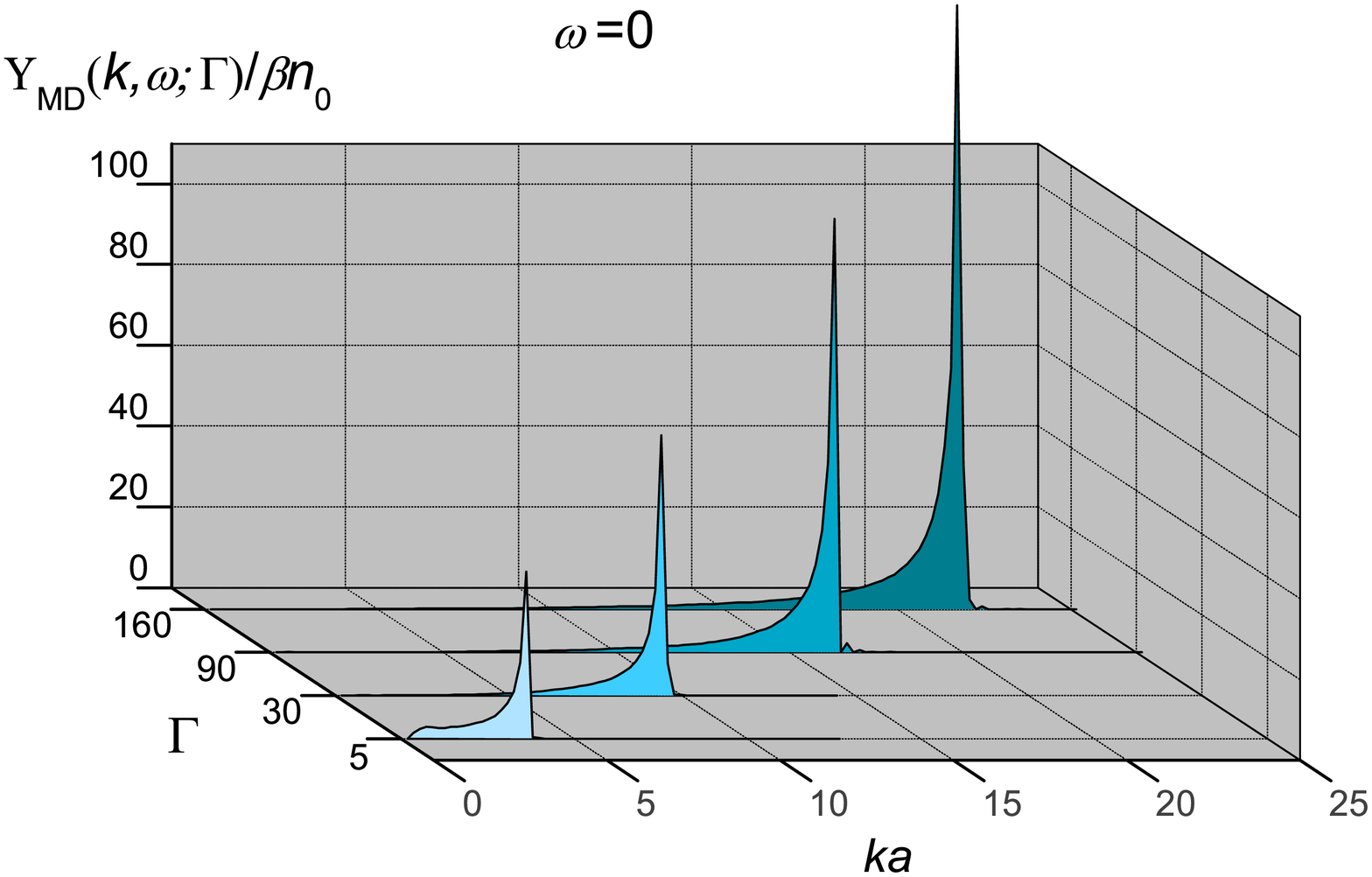}
\includegraphics[width=1.0\columnwidth]{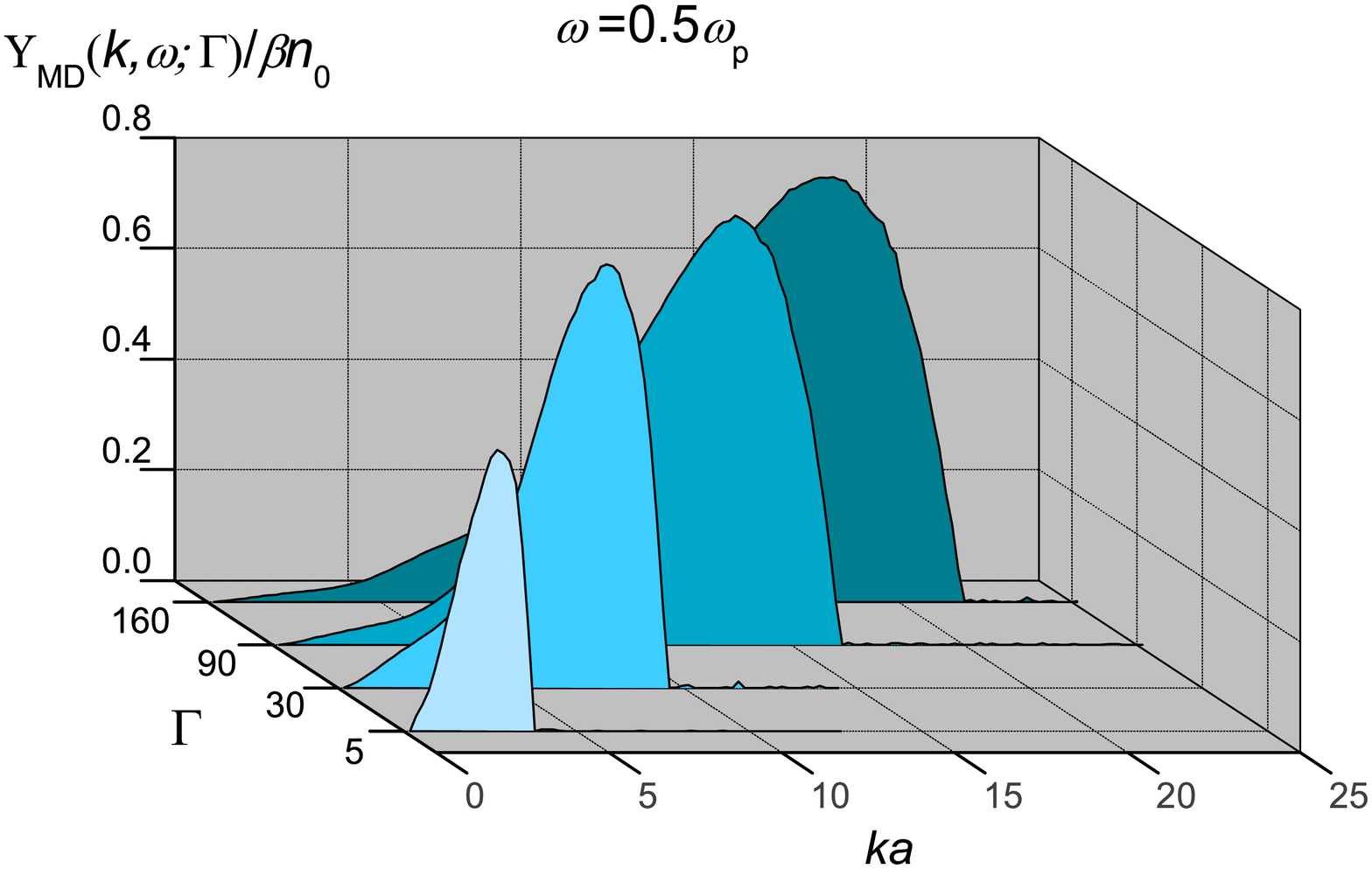}
\caption{The behavior of the violating (anomalous) term $\Upsilon_{\rm MD}(\textbf{k},\omega)$ as a function of $k$ at $\omega=0$ and $\omega=0.5\,\omega_{\rm p}$ for the range of $\Gamma$ values indicated. Observe that the violation disappears for $k>k_\ast(\Gamma)$.}
\label{Upsilon-w}
\end{figure}

\begin{figure}[htb]
\includegraphics[width=1.0\columnwidth]{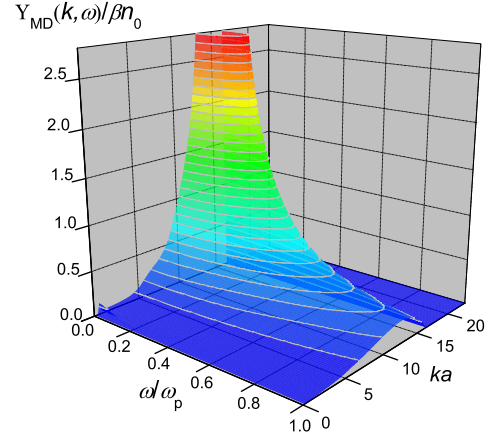}
\caption{$\Upsilon_{\rm MD}(\textbf{k},\omega)$ as a function of $k$ and $\omega$ at $\Gamma=90$.}
\label{Upsilon-surface}
\end{figure}

Within the $0<k<k_{\ast}$ domain the violation can be characterized by the difference
\begin{equation}
\Upsilon_{\rm MD}(\textbf{k},\omega)=\overline{\chi}'(\textbf{k},\omega)-\frac{2}{\pi}
\mathcal{P}\int_{0}^{\infty}\nu\frac{\overline{\chi}''(\textbf{k},\nu)}{\nu^{2}-\omega^{2}}d\nu.
\label{}
\end{equation}
The explicit determination of this violating term (the MD subscript refers to the determination of this quantity from the computed values of $\overline{\chi}'(\textbf{k},\omega)$ and $\overline{\chi}''(\textbf{k},\omega)$), which we will refer to as its $"anomalous"$ part is one of the main results of the present work. The behavior of $\Upsilon_{\rm MD}(\textbf{k},\omega)$ as a function of $\textbf{k}$ and $\omega$ for a range of $\Gamma$ values is displayed in Fig.~\ref{Upsilon-w}. The static $\Upsilon_{\rm MD}(\textbf{k})$ exhibits a singularity at $k=k_\ast$, which, however, is removed at finite frequencies. The full $\Upsilon_{\rm MD}(\textbf{k},\omega)$ landscape in Fig.~\ref{Upsilon-surface}. is compatible with the behavior that the violation extends to $\omega\rightarrow\infty$ (see below). As to the $\Gamma$ dependence, it can be seen that stronger coupling generates more substantial violation.


The key features of the anomaly revealed by the MD simulation  can be understood on an analytic basis.

$\overline{\chi}(\textbf{k},\omega)$ not being an analytic function on the upper half plane, it must have singularities there. In fact, it has rigorously been shown by Losyakov \cite{L} that it has one single simple pole on the imaginary axis, say at $iy(\textbf{k};\Gamma)$. In order to maintain the required symmetry of the response, it must be accompanied by an other pole at $-iy(\textbf{k};\Gamma)$. If the residues at the poles are
\begin{equation}
Res_{\omega=\pm iy}\overline{\chi}(\textbf{k},\omega)=\mp\frac{i}{2}Q(\textbf{k};\Gamma),
\label{}
\end{equation}
then the contribution to $\overline{\chi}(\textbf{k},\omega)$ from these singularities on the real $\omega$-axis becomes the real function
\begin{equation}
\Upsilon(\textbf{k},\omega)=\frac{Q(\textbf{k}) y(\textbf{k})}{\omega^{2}+y^{2}(\textbf{k})}.
\label{Upsilon}
\end{equation}

We can refer to $\Upsilon(\textbf{k},\omega)$ as the $"anomalous"$ part of the response (the structure of which  shows that the violation extends to $\omega\rightarrow\infty$). Once subtracted from $\overline{\chi}(\textbf{k},\omega)$
the remaining
\begin{equation}
\Xi(\textbf{k},\omega)=\overline{\chi}(\textbf{k},\omega)-\Upsilon(\textbf{k},\omega)
\label{}
\end{equation}
is the $"regular"$ part of the response. This latter now is a $plus$-function and it satisfies the KK relations. The first of these relations can be written as
\begin{align}
\Xi'(\textbf{k},\omega)
&=
\overline{\chi}'(\textbf{k},\omega)-\Upsilon(\textbf{k},\omega) \nonumber \\
&=\frac{2}{\pi}\mathcal{P}\int_{0}^{\infty}
\nu\frac{\Xi''(\textbf{k},\nu)}{\nu^{2}-\omega^{2}}d\nu \nonumber \\
&= \frac{2}{\pi}\mathcal{P}\int_{0}^{\infty}
\nu\frac{\overline{\chi}''(\textbf{k},\nu)}{\nu^{2}-\omega^{2}}d\nu.
\label{mKK1}
\end{align}
Introducing now the Hilbert transform
\begin{equation}
\Lambda(\textbf{k},\omega)=-\frac{i}{\pi}\mathcal{P}\int_{-\infty}^{\infty}
\frac{\overline{\chi}(\textbf{k},\nu)}{\nu-\omega}d\nu,
\end{equation}
the above equation may be cast in the form
\begin{equation}
\overline{\chi}'(\textbf{k},\omega)=  \Lambda'(\textbf{k},\omega)+\Upsilon(\textbf{k},\omega).
\label{mKK1-lambda}
\end{equation}

\begin{figure}[htb]
\includegraphics[width=1.0\columnwidth]{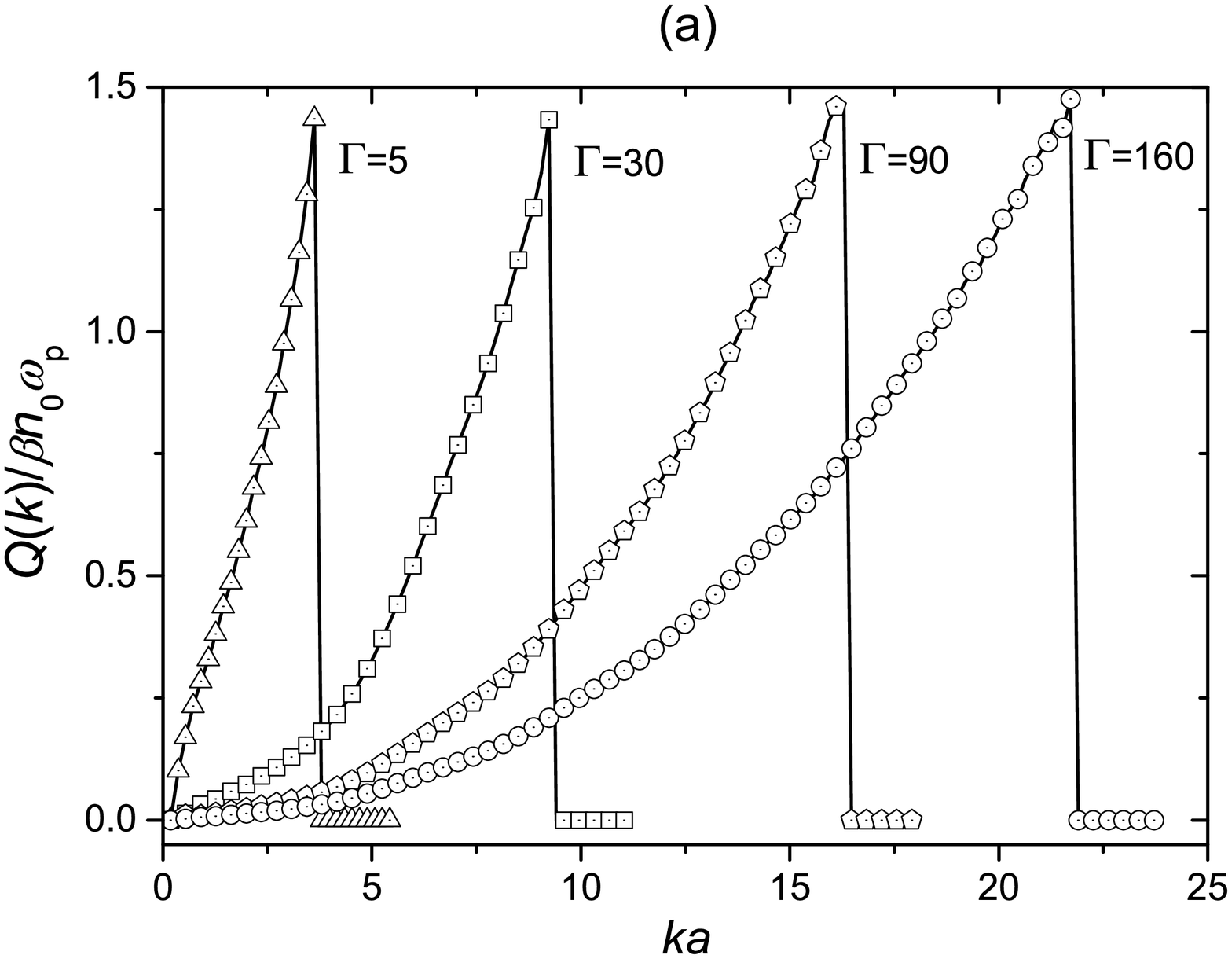}
\includegraphics[width=1.0\columnwidth]{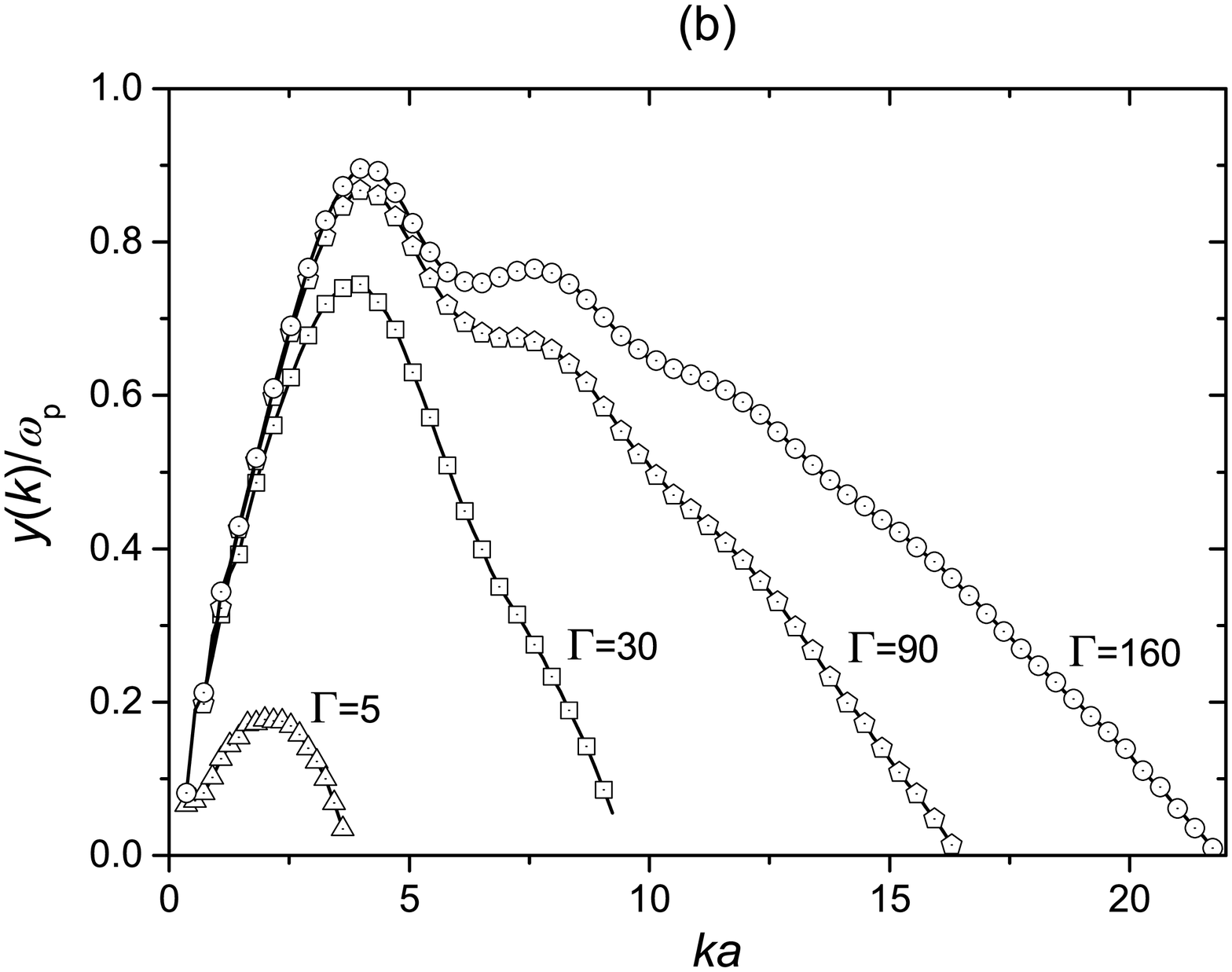}
\caption{Dependence of the $Q(\textbf{k})$ and $y(\textbf{k})$ parameters of the imaginary pole on the wave-number for $\Gamma$ values indicated.}
\label{vio-parameters}
\end{figure}

\begin{figure}[htb]
\includegraphics[width=1.0\columnwidth]{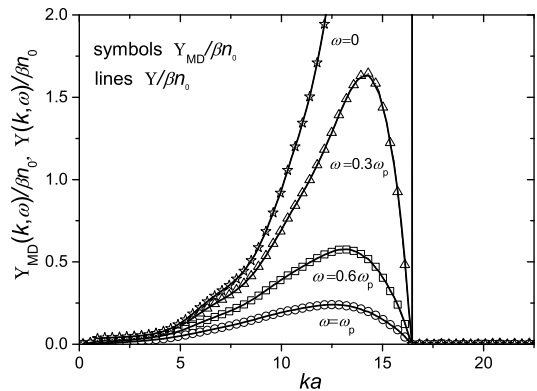}
\caption{$\Upsilon(\textbf{k},\omega)$ obtained from the computed values of $Q$ and $y$ parameters, in comparison with the MD generated $\Upsilon_{\rm MD}(\textbf{k},\omega)$ for $\Gamma=90$.}
\label{Upsilon-compare}
\end{figure}

Based on the above relationship we can now proceed to determine the two parameters $y(\textbf{k})$, $Q(\textbf{k})$ by re-defining the  sum rules that the physical response $\overline{\chi}(\textbf{k},\omega)$ now satisfies. Setting first $\omega=0$  we obtain  the extended thermodynamic sum rule
\begin{equation}
\frac{Q(\textbf{k})}{y(\textbf{k})}=-\frac{2}{\pi}\mathcal{P}\int_{0}^{\infty}
\frac{\overline{\chi}''(\textbf{k},\nu)}{\nu}d\nu+\overline{\chi}'(\textbf{k},0).
\label{Qy1}
\end{equation}
Letting $\omega\to\infty$ the requirement that $\overline{\chi}'(\textbf{k},\omega\rightarrow\infty)=\frac{n_0}{m}\frac{k^{2}}{\omega^{2}}$ yields the extended $f$-sum rule
\begin{equation}
Q(\textbf{k})y(\textbf{k})=\frac{n_0}{m}k^{2}+\frac{2}{\pi}\int_{0}^{\infty}\nu\overline{\chi}''(\textbf{k},\nu)d\nu.
\label{Qy2}
\end{equation}
Fig.~\ref{vio-parameters} shows $Q(\textbf{k};\Gamma)$ and $y(\textbf{k};\Gamma)$ as calculated from Eqs. (\ref{Qy1}) and (\ref{Qy2}) using the MD data. Equipped with this information, we are now able to generate $\Upsilon(\textbf{k},\omega)$ and compare it with the MD generated $\Upsilon_{\rm MD}(\textbf{k},\omega)$ obtained before. This is done for $\Gamma=90$ in Fig.~\ref{Upsilon-compare}. The agreement is excellent, verifying the reliability of the MD protocol and the soundness of the analysis.

Considering the implications of the second KK relationship for
$\Xi''(\textbf{k},\omega)$:
\begin{align}
\Xi''(\textbf{k},\omega)&=
\overline{\chi}''(\textbf{k},\omega) \nonumber \\
&=-\frac{2}{\pi}\omega\mathcal{P}\int_{0}^{\infty}
\frac{\overline{\chi}'(\textbf{k},\nu)}{\nu^{2}-\omega^{2}}d\nu \nonumber \\
&\quad+\frac{2}{\pi}\omega\mathcal{P}\int_{0}^{\infty}\frac{1}{\nu^{2}-\omega^{2}}\frac{Q(\textbf{k})y(\textbf{k})}{\nu^2+y^{2}(\textbf{k})}d\nu \nonumber \\
&=  \Lambda''(\textbf{k},\omega)
-\frac{\omega}{y(\textbf{k})}\Upsilon(\textbf{k},\omega),
\label{mKK2}
\end{align}
we observe that even though $\Upsilon(\textbf{k},\omega)$
is a real function and therefore has no contribution to the imaginary part of $\overline{\chi}(\textbf{k},\omega)$,
when the latter is expressed via the Hilbert transform of $\overline{\chi}'(\textbf{k},\omega)$, $\Upsilon(\textbf{k},\omega)$
provides a complementary term, mirroring the architecture of Eq. (\ref{mKK1-lambda}).\par

In the Eq. (\ref{mKK2}) only the $\omega\rightarrow\infty$ limit gives a useful result. Asserting that $ \overline{\chi}''(\textbf{k},\omega)$ vanishes faster than $1/\omega$ as $\omega\rightarrow\infty$, one finds
\begin{equation}
\frac{2}{\pi}\int_{0}^{\infty}
\overline{\chi}'(\textbf{k},\nu)d\nu
=Q(\textbf{k}),
\end{equation}
and we arrive at the extension of one of the ADNS sum rule \cite{ADNS}.
This relationship is redundant, but serves as a useful consistency check. \par
\begin{figure}[htb]
\includegraphics[width=1.0\columnwidth]{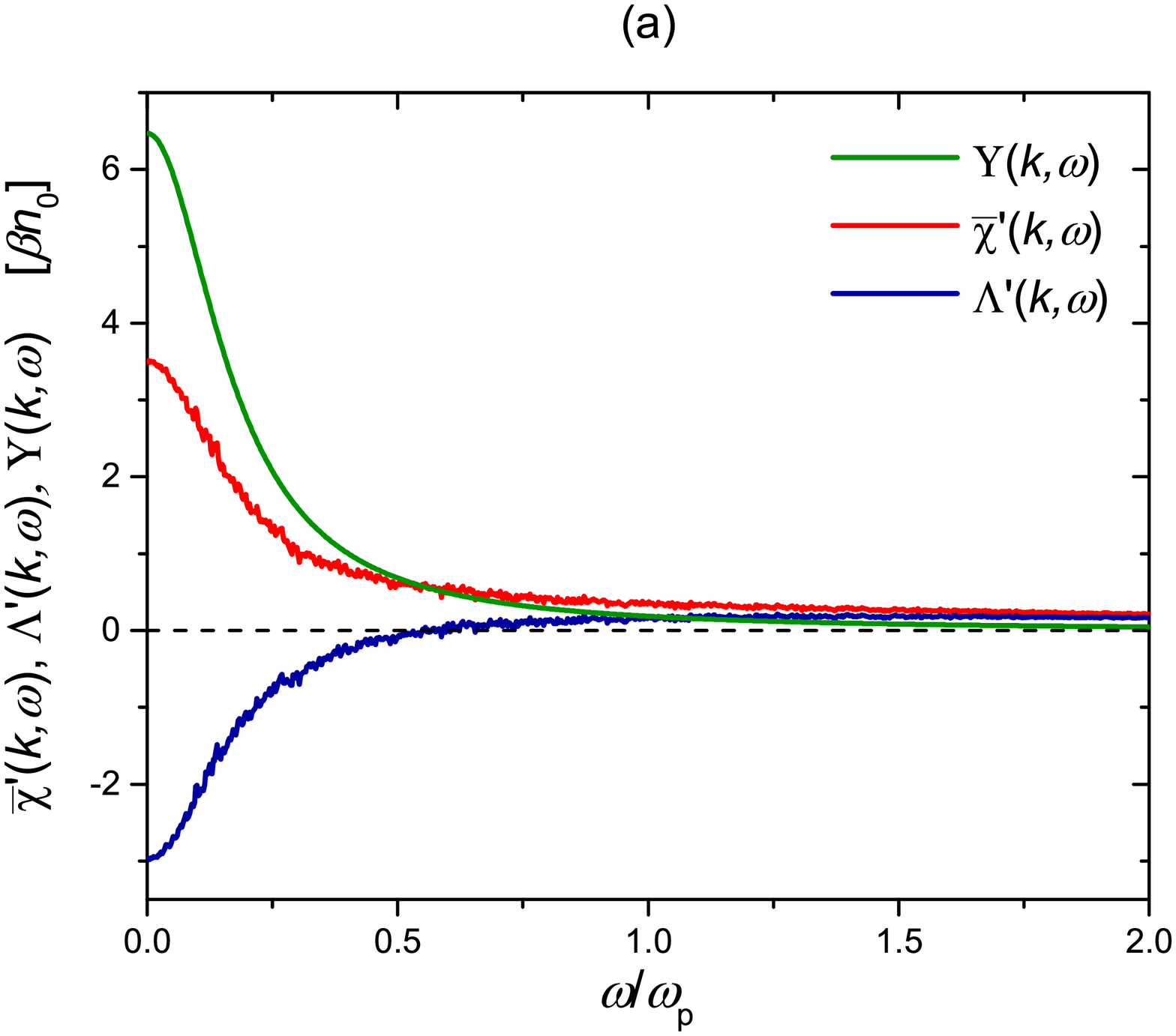}
\includegraphics[width=1.0\columnwidth]{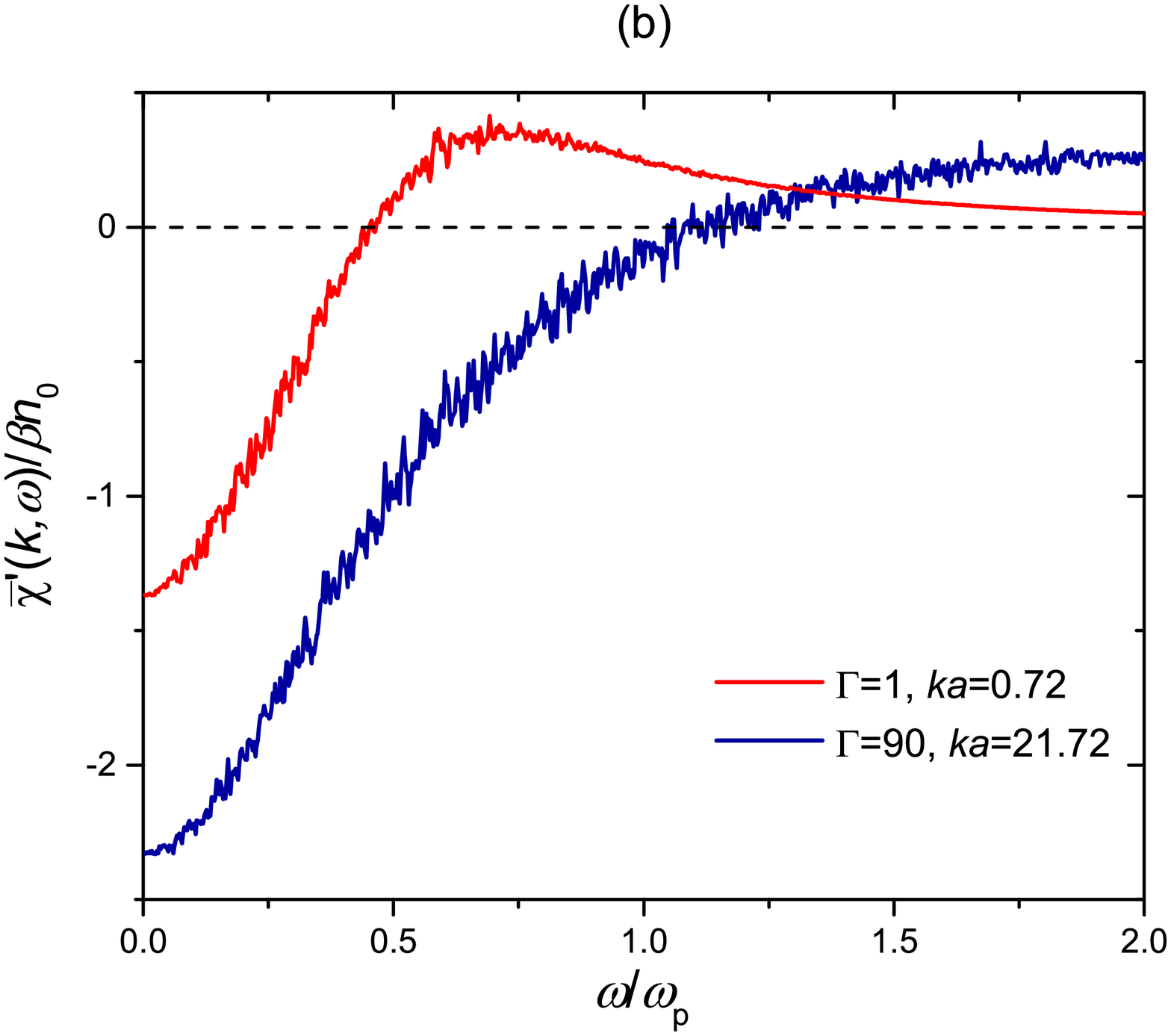}
\caption{(a) Two contributions to the  real part of the response function $\overline{\chi}'(\textbf{k},\omega)$ in the domain of the violation (here, at $\Gamma=90$ and $ka=14.48<k_{\ast}a=16.38$):  the real part of the Hilbert transform, $\Lambda'(\textbf{k},\omega)$ and the anomalous $\Upsilon(\textbf{k},\omega)$ term, corresponding to the modified KK relation (\ref{mKK1-lambda}). (b) Real part of the response function $\overline{\chi}'(\textbf{k},\omega)$ outside of the domain of the violation, at the $\Gamma$ and $ka$ values indicated.}
\label{Rechitot}
\end{figure}
\begin{figure}[htb]
\includegraphics[width=1.0\columnwidth]{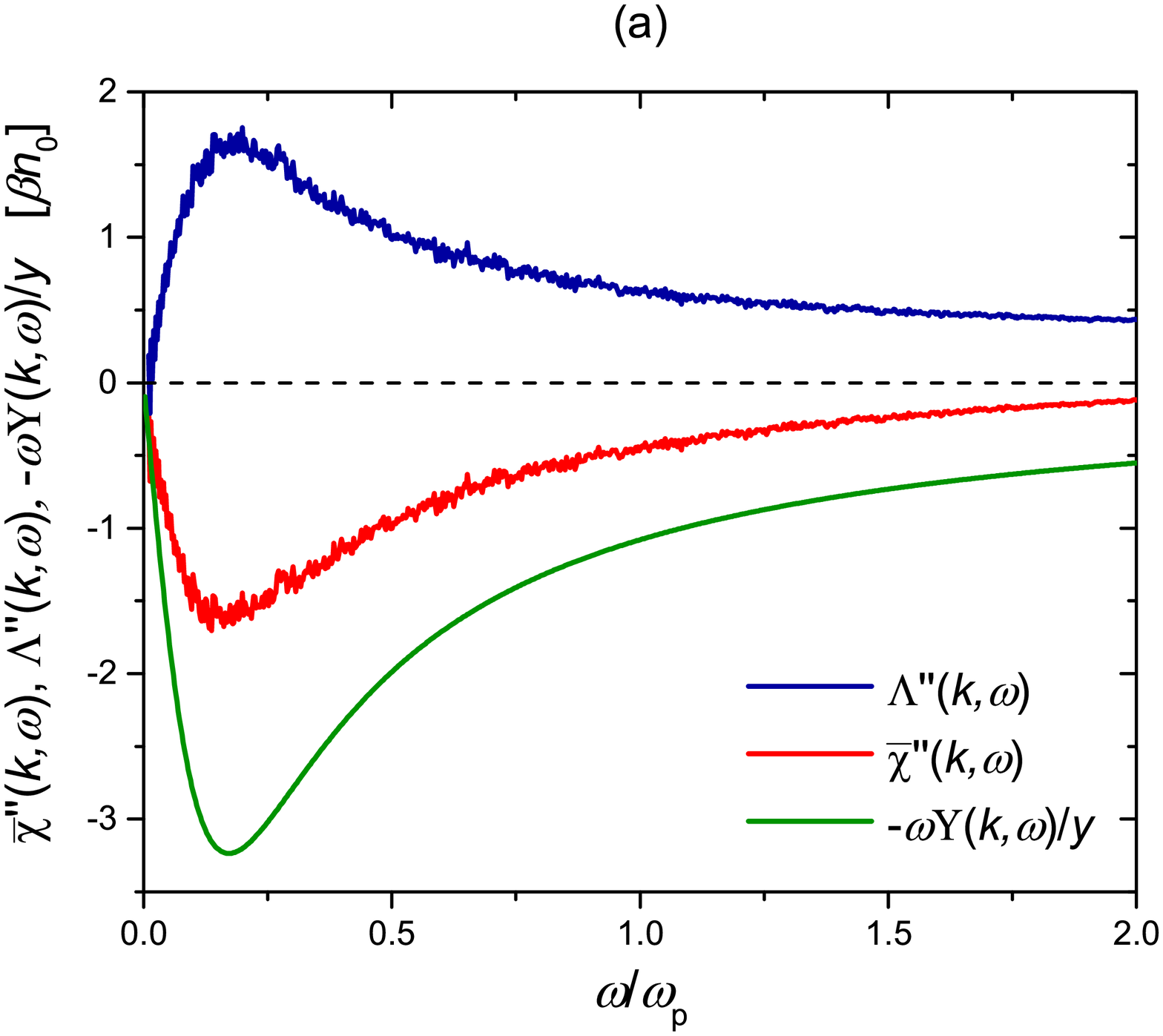}
\includegraphics[width=1.0\columnwidth]{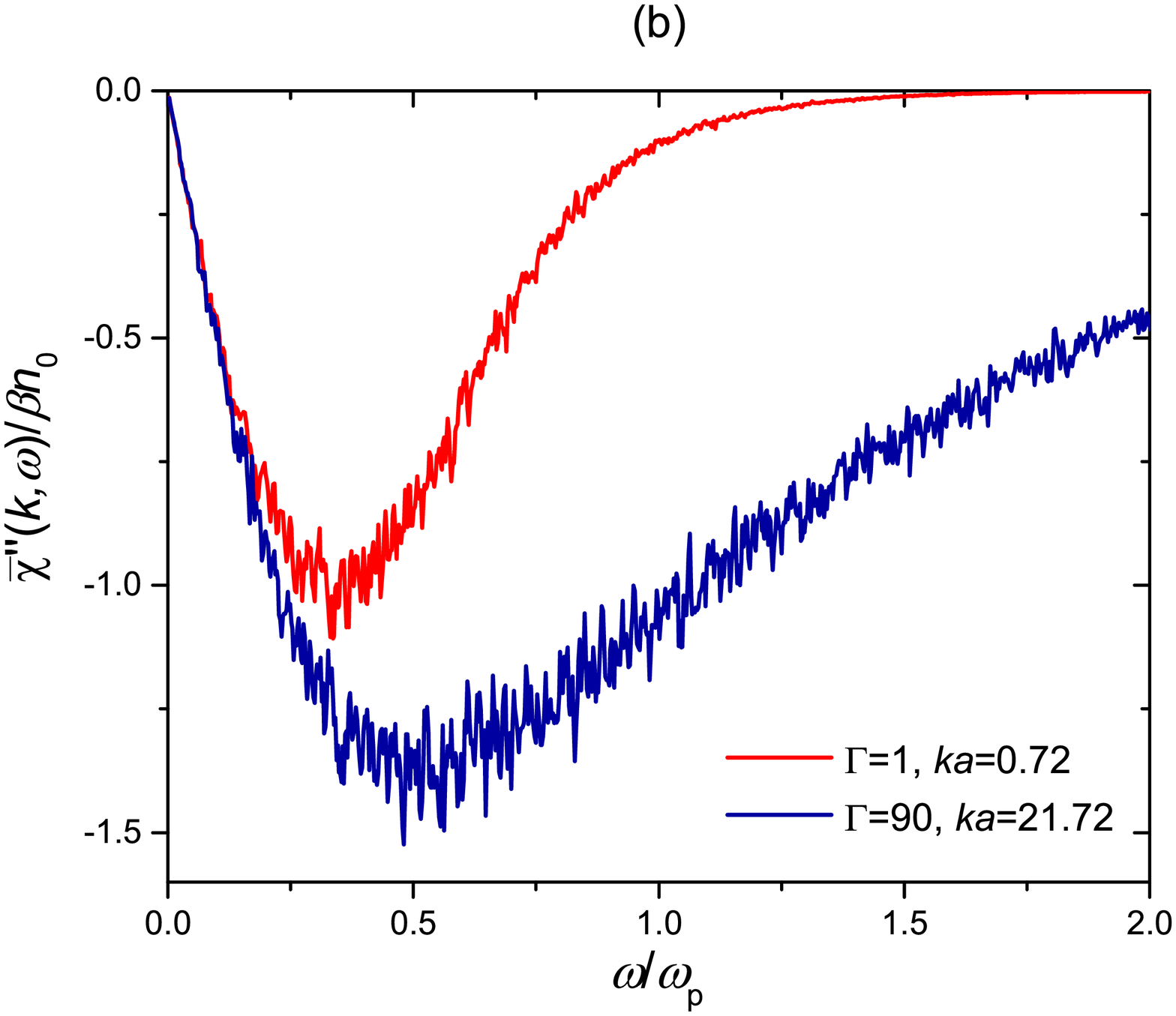}
\caption{(a) Two contributions to the  imaginary part of the response function $\overline{\chi}''(\textbf{k},\omega)$ in the domain of the violation (here, at $\Gamma=90$ and $ka=14.48<k_{\ast}a=16.38$):  the imaginary part of the Hilbert transform, $\Lambda''(\textbf{k},\omega)$ and the anomalous $-\omega\Upsilon(\textbf{k},\omega)/y(\textbf{k})$ term, corresponding to the modified KK relation (\ref{mKK2}). (b) Imaginary part of the response function $\overline{\chi}''(\textbf{k},\omega)$ outside of the domain of the violation, at the $\Gamma$ and $ka$ values indicated.}
\label{Imchitot}
\end{figure}

To summarize the results of this Section, in Figs.~\ref{Rechitot} and \ref{Imchitot} we present a series of graphs showing the real and imaginary parts of $\overline{\chi}(\textbf{k},\omega)$, both in the domains of normal behavior ($\Gamma<\Gamma_{\ast}$ or $k>k_{\ast}$) and of the violation. In the latter, we also show the split into the Hilbert transform $\Lambda(\textbf{k},\omega)$  and anomalous $\Upsilon(\textbf{k},\omega)$
contributions. It is interesting to note  that in this domain both the real and imaginary parts of $\overline{\chi}(\textbf{k},\omega)$
are dominated by their anomalous parts. \\   \par
\begin{figure}[htb]
\includegraphics[width=1.0\columnwidth]{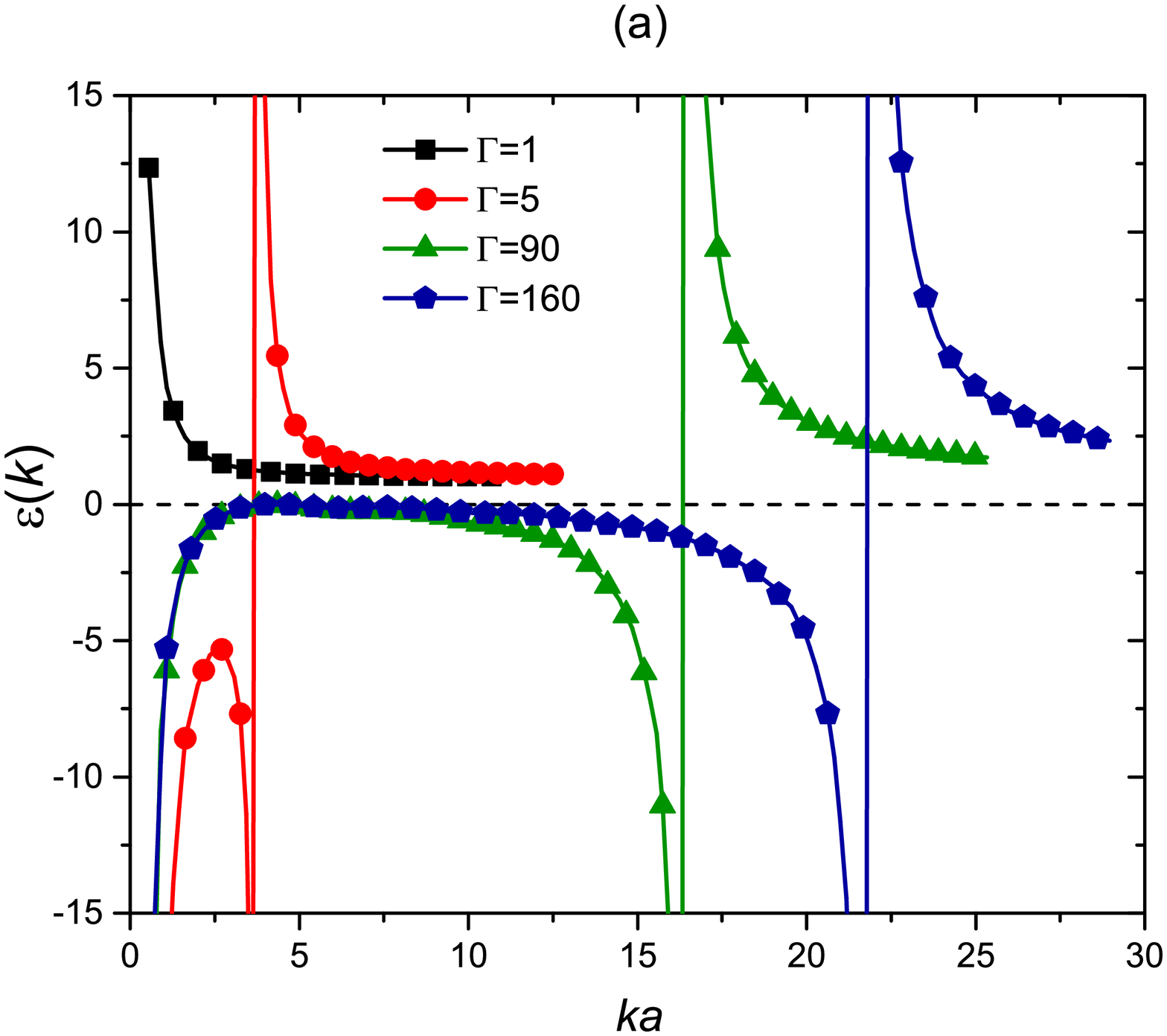}
\includegraphics[width=1.0\columnwidth]{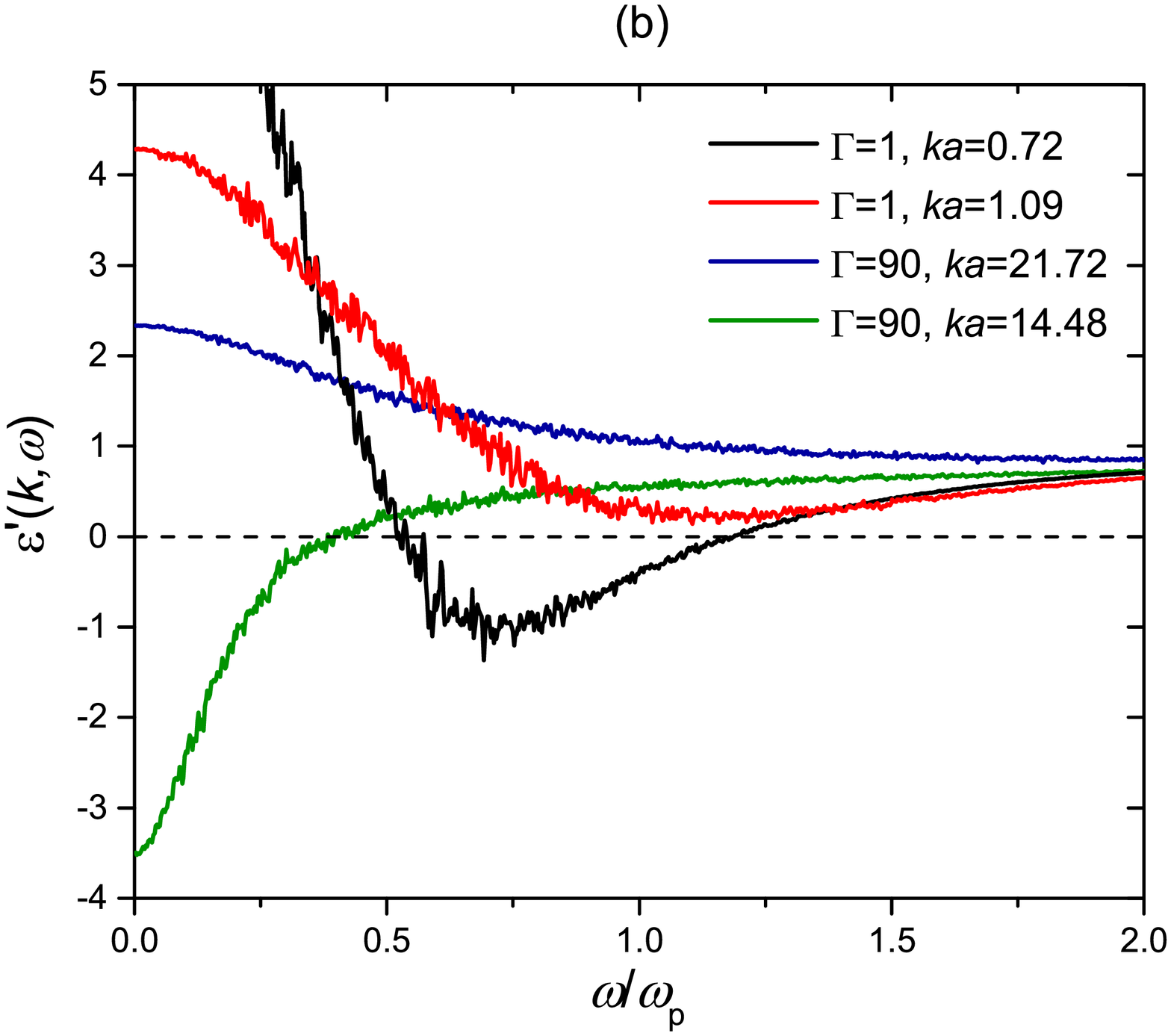}
\caption{(a) Static dielectric response function $\varepsilon(\textbf{k})$ for $\Gamma$ values indicated. (b) Dependence of the real part of the dielectric response function $\varepsilon'(\textbf{k},\omega)$ on the frequency at the $\Gamma$ and $ka$ values indicated. We note that when $\Gamma>\Gamma_\ast$ the intersection of $\varepsilon'(\textbf{k},\omega)$ with the zero axis disappears for $k>k_\ast$.}
\label{Reeps}
\end{figure}

Finally, for the sake of completeness in Figs.~\ref{Reeps} we also show the static $\varepsilon(\textbf{k})$ and the frequency dependence of the dielectric function. $\varepsilon(\textbf{k},\omega)$, in particular $\varepsilon(\textbf{k})$ becomes negative along $\overline{\chi}'(\textbf{k},\omega)$, in the violating domain. The latter exhibits the characteristic inverted U-shape, predicted by Kirzhnits. It also satisfies the stability criterion  $\varepsilon (\bf k )>$ 1 in the normal and $\varepsilon (\bf k )<$ 0 in the violating region \cite{Keldysh}. The role of a negative static dielectric function in the formation of the  ground state of the electron liquid has recently drawn attention in the literature \cite{Schakel}.\par

\section{Collective excitation}
The question how the negative compressibility and the concomitant acausal behavior  affect the behavior of the plasmon, the collective excitation of  the OCP, is not quite meaningful, since negative compressibility is only one of the various consequences of strong coupling and what is open to observation is only the combined effect of all these components. Inferences, however, can be drawn, primarily by observing the role played by the $y-pole$, the hallmark of acausal behavior. We now contend that there are at least  two distinct patterns of behavior in the plasmon dispersion, which can be reasonably attributed to the existence of the $y-pole$.\par

Recalling that $\overline{\chi}(\textbf{k},\omega)$
consists of a regular and an anomalous part, we observe from Figs. 2(b) and 7(a) that at high $\Gamma$ values, and especially in the high $\omega$ domain,  the anomalous $\Upsilon(\textbf{k},\omega)$
part becomes dominant. To asses the influence of this anomalous part on the plasmon dispersion we propose to create a truncated dielectric response function
\begin{equation}
\widetilde{\varepsilon}(\textbf{k},\omega)=1-\varphi(\textbf{k})\Upsilon(\textbf{k},\omega)
\label{tr-eps}
\end{equation}
and examine the approximate dispersion relation
\begin{equation}
\widetilde{\varepsilon}\,'(\textbf{k},\omega)=0
\label{tr-dispersion}
\end{equation}

\begin{figure}[htb]
\includegraphics[width=1.0\columnwidth]{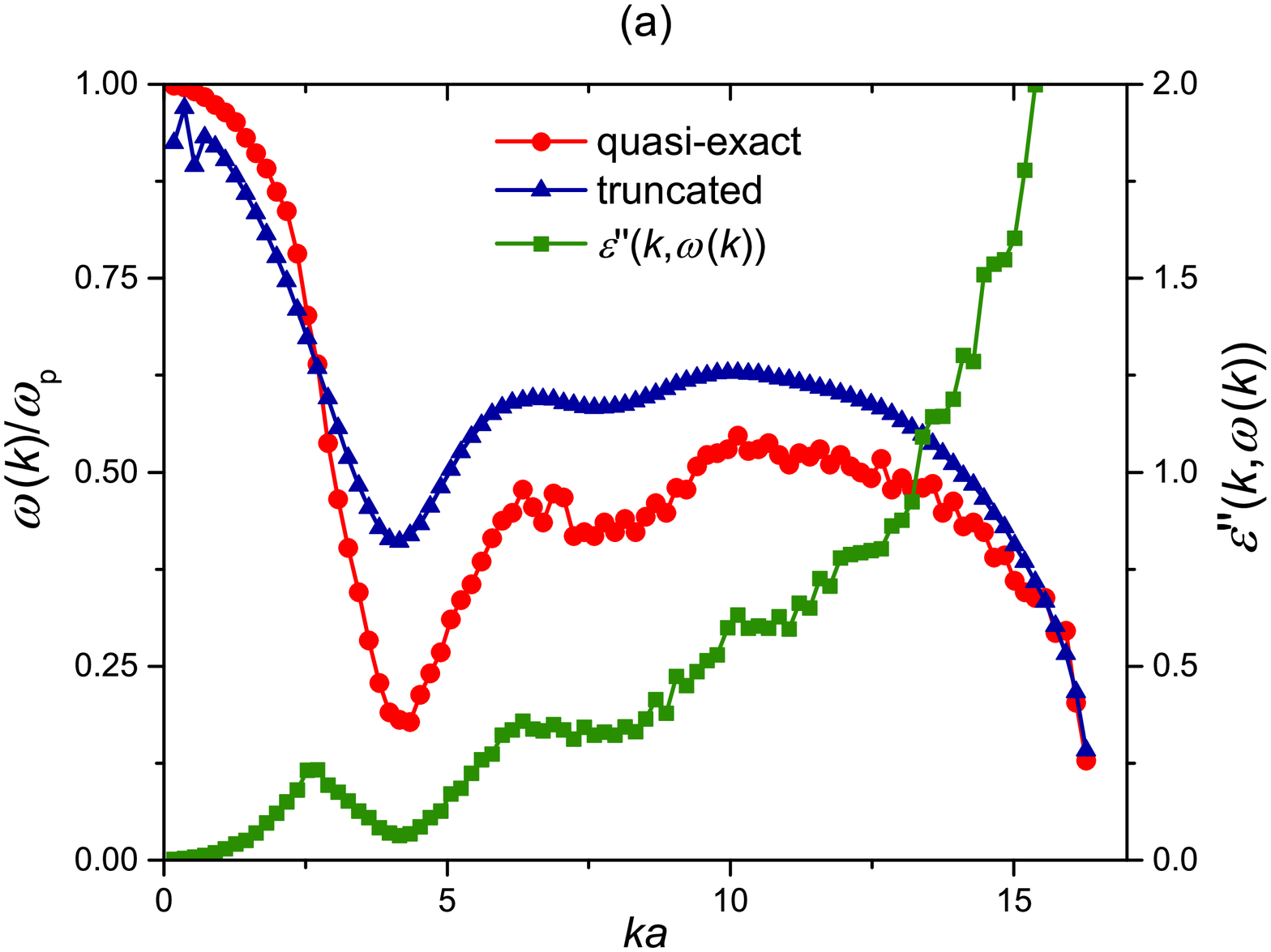}
\includegraphics[width=1.0\columnwidth]{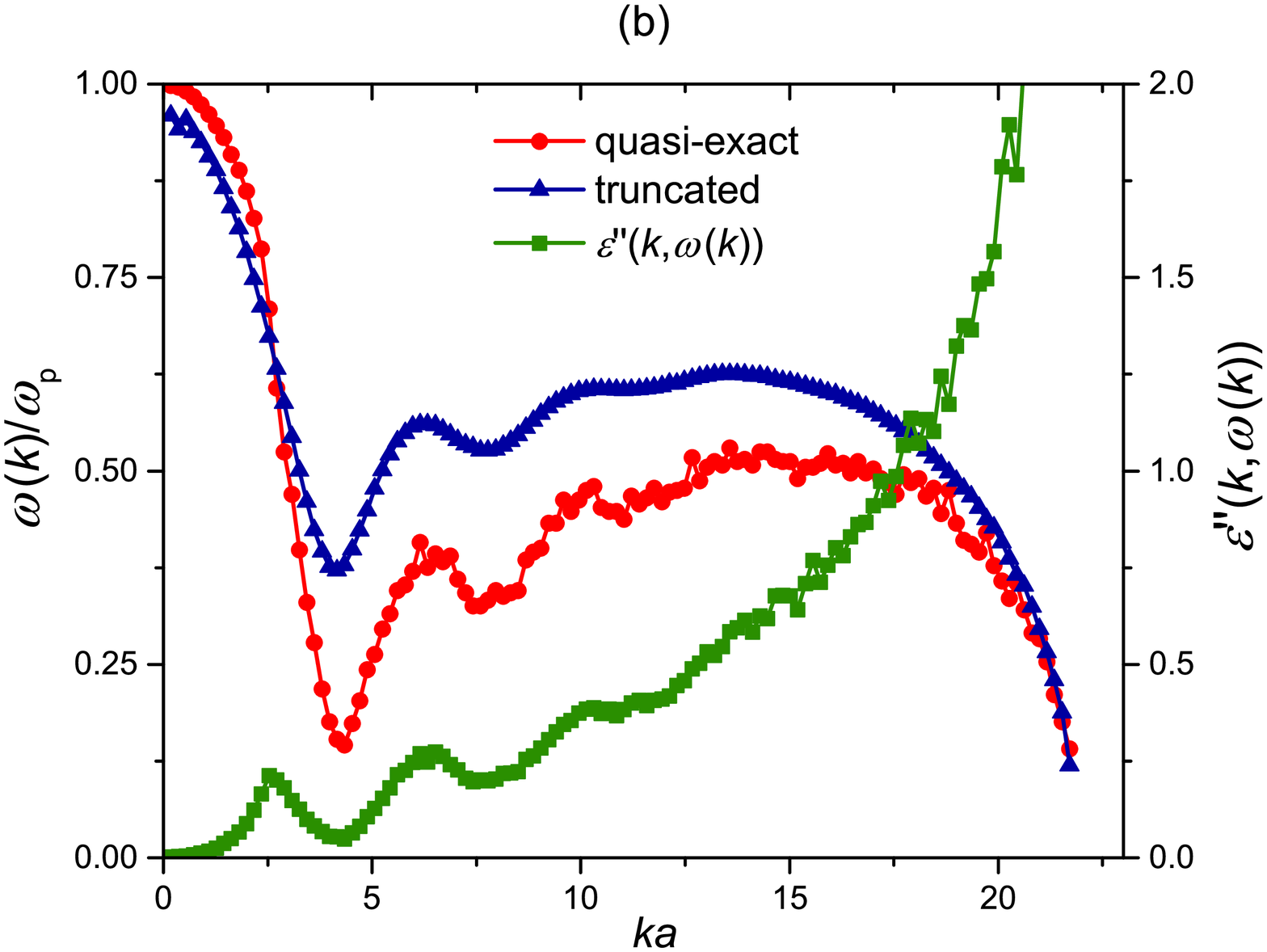}
\caption{The truncated dispersion relation resulting from $\widetilde{\varepsilon}\,'(\textbf{k},\omega)=0$, in comparison with the quasi-exact relation obtained from $\varepsilon'(\textbf{k},\omega)=0$. The dispersions terminate at $\omega_{\ast}=\omega(k_{\ast})$.
The imaginary part of the dielectric response function at the quasi-exact dispersion frequency is also shown. (a) $\Gamma=90$, (b) $\Gamma=160$.}
\label{dispersion}
\end{figure}

The result of this procedure compared with the quasi-exact (the qualifier refers to the neglect of $\varepsilon''(\textbf{k},\omega)$) plasmon dispersion calculated from
\begin{equation}
\varepsilon'(\textbf{k},\omega)=0
\label{eq:epsprime}
\end{equation}
is shown in Fig.~\ref{dispersion}. A major feature of the  plasmon dispersion at strong coupling is the formation of the so-called $roton-minimum$ for $k_{\rm RM}a=\overline{k}_{\rm RM}\simeq4.5$ at $\omega_{\rm RM}/\omega_{\rm p}\simeq0.25$ \cite{Ihor}. The roton minimum was originally identified in the collective mode spectrum of liquid He \cite{Landau,Griffin}, (according to the most recent measurements \cite{He4} the position of the roton minimum in $^{4}{\rm He}$ is at $\overline{k}=4.3$), but by now it is fairly well understood that it is the common feature of most strongly correlated many-body systems \cite{Kyrkos,Nozieres}. What we observe now is that the truncated dispersion relation, Eq.(\ref{tr-dispersion}) reproduces the roton minimum of the quasi-exact spectrum with remarkable accuracy. In particular, the position of the minimum is determined by the position of the maximum of $y(k)$. In other words, the roton minimum is the consequence of the presence of the $y-pole$ in the response function. Then the conclusion, which is certainly valid for the OCP studied here, but may  reasonably be surmised to be of more general applicability, that the roton minimum is the consequence of the negative compressibility of the system, follows. \par

At the $k=k_{\ast}$ singularity $\varepsilon(\textbf{k})$ changes sign
and remains positive for all wave numbers (see Fig 8). Consequently, in this domain Eq. (\ref{eq:epsprime}) ceases to lead to real $\omega$ solutions. The dispersion curve seems to terminate at a finite $\omega_{\ast}(\Gamma)=\omega(k_{\ast}(\Gamma))$ frequency, which seems by inspection to be the lowest frequency the system can reach in the liquid phase. However, the limited resolution of the MD simulation doesn't allow us to state with certainty that the dispersion curve does not continue all the way down to $\omega=0$. In either case, the high value of damping in this high $k$-domain (also shown in Fig. 9) makes these nominal frequency values of little physical significance. How these results can be reconciled with the result of the QLCA analysis \cite{QLCA,Yukawa-1,Yukawa-2,Yukawa-3,Trilayer} that predicts that in general for any Coulomb-like system $\omega(k\to\infty)=\Omega_{\rm E}$, the Einstein frequency of the system (and with the  verifying   MD and experimental \cite{Nunomura} findings) will be discussed elsewhere (see also a recent discussion on the various possible approaches to and interpretations of the plasmon dispersion in \cite{Vorberger}).

\begin{figure}[htb]
\includegraphics[width=1.0\columnwidth]{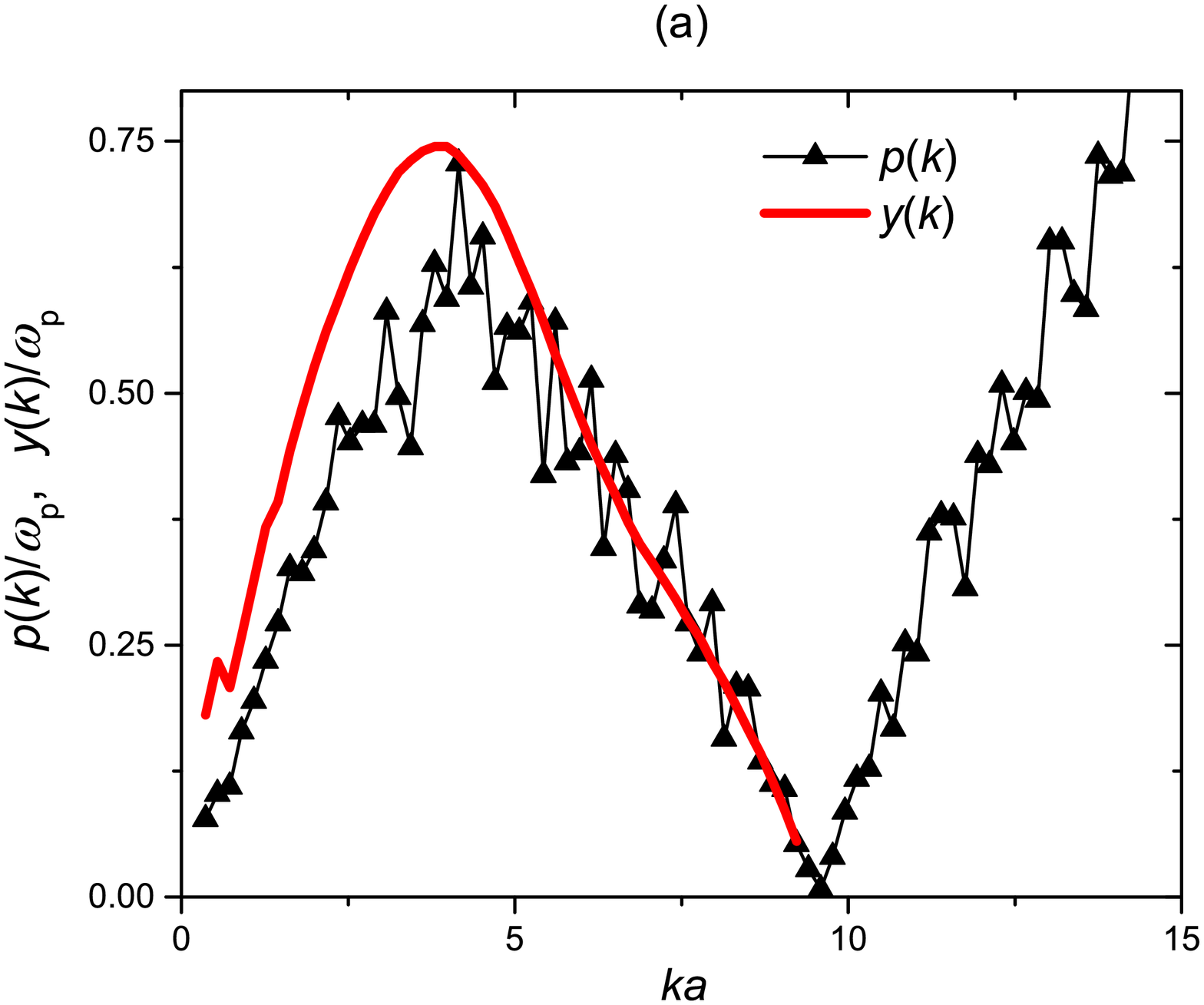}
\includegraphics[width=1.0\columnwidth]{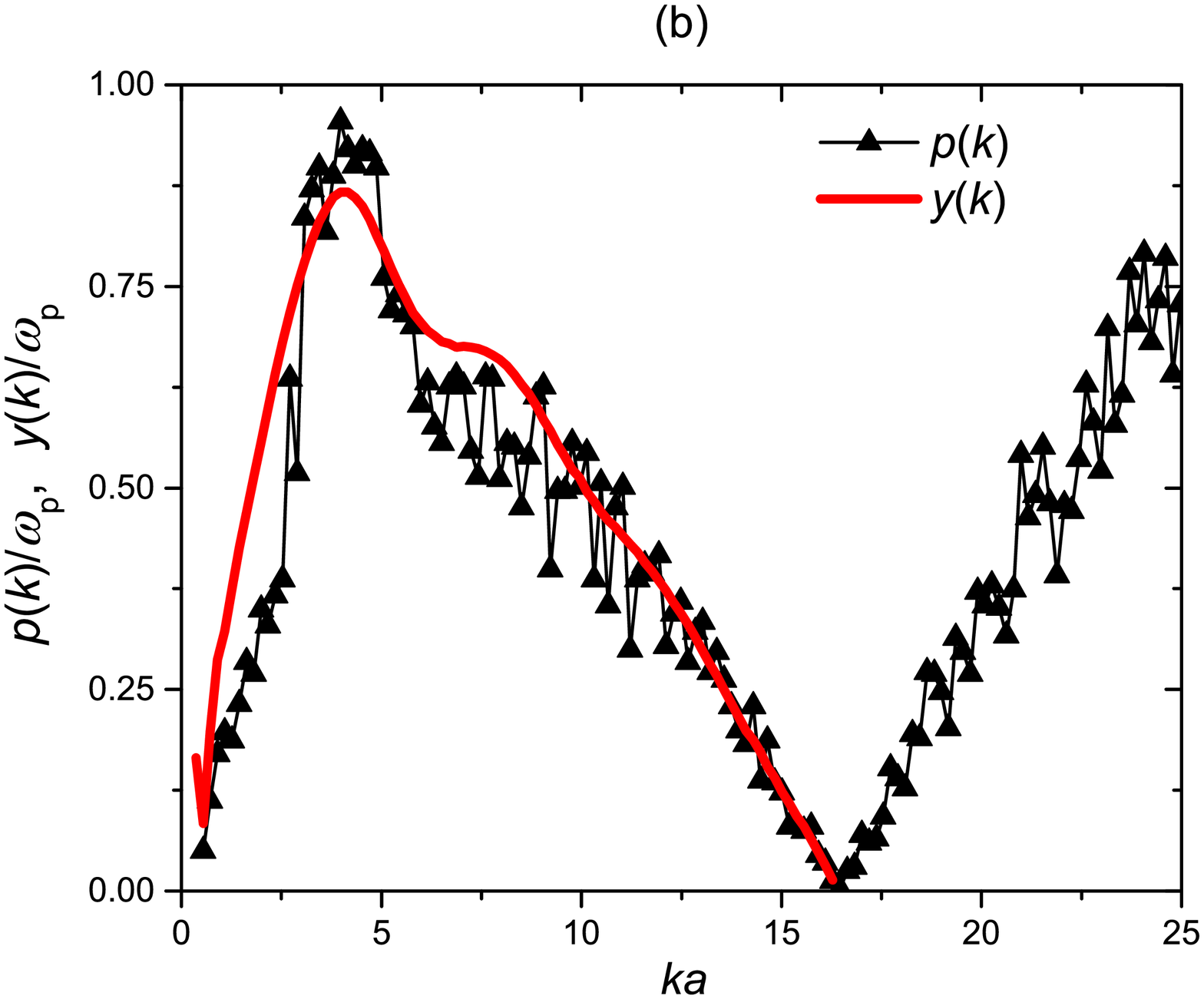}
\caption{Comparison of the $\textbf{k}$-dependence of the peak-position $p(\textbf{k})$ of $\overline{\chi}''(\textbf{k},\omega)$ with the $\textbf{k}$-dependence of $y(\textbf{k})$ parameter: at (a) $\Gamma=30$ and (b) $\Gamma=90$.}
\label{peak}
\end{figure}

We now turn to examining the influence of the $y-pole$ on $\varepsilon''(\textbf{k},\omega)=-\varphi(\textbf{k})\overline{\chi}''(\textbf{k},\omega)$, which is responsible for generating the damping of plasmons. Our detailed study of $\overline{\chi}''(\textbf{k},\omega)$ across coupling domains \cite{Future}, especially its $\omega$-dependence, verifies that the shape of the function, which for $\Gamma\to 0$ in the RPA emulates the derivative of the Maxwell velocity distribution function, remains grossly invariant under $\Gamma$ being increased to higher values (cf. Fig.~\ref{Imchitot}(b)). Thus its behavior is well characterized by the two parameters $p(\textbf{k})$ and $h(\textbf{k})$, the position and height of its peak value. Characteristically, for $\Gamma<\Gamma_{\ast}$, $p(\textbf{k})/\overline{k}$ stays in the vicinity of $p(\textbf{k})/\overline{k}=b(\Gamma)$, a $\textbf{k}$-independent constant. However, as demonstrated by Fig.~\ref{peak}, a dramatic change in this behavior occurs once one is inside the violation domain: here $p(\textbf{k})$ closely follows the non-monotonic $k$-dependence of $y(\textbf{k})$: $p(\textbf{k})\simeq y(\textbf{k})$. While we do not have a clear understanding of the physics that brings about this feature, a brief model calculation in the next Section provides an insight of how the interaction between the dissipation and the $y-pole$ leads to this peculiar phenomenon.\par
We also note that the damping of the oscillation at the roton minimum is quite low \cite{Ihor}, ensuring that it is a well-defined collective excitation. The reason for this can be understood on the basis of  what has been  discussed in the previous paragraph. Since the damping is determined by the value of $\overline{\chi}''(k_{\rm RM},\omega_{\rm RM})$, a little reflection shows that the behavior described above maximizes the separation between $y$ and $\omega_{\rm RM}$, thereby forcing the latter into the tail of the distribution, where $\overline\chi''$ assumes only  a very low value.

\section{Model Calculation}
The foregoing derivation provides a completely satisfactory  formal description of the acausal behavior of the class of response functions in question. Nevertheless, this behavior is sufficiently lacking in intuitive appeal to make it  desirable to seek a physically more understandable explanation of its origin.

While it is clear from the integral (\ref{integral}) and the discussion leading to it that it is the negative compressibility of the system that ultimately causes the breakdown of causal behavior, what is intuitively certainly not clear, indeed somewhat mysterious, how these two, seemingly physically unrelated phenomena are linked with each other. To see this, we examine a simple two-pole model with a $\textbf{k}$-dependent collision frequency or damping rate, $\gamma(\textbf{k})$ with $\gamma(k\rightarrow 0)=0$. In this approximation the external density response function takes the form
\begin{equation}
\chi(\textbf{k},\omega)=\frac{A}{2}\left[\frac{1}{\omega-(\omega(\textbf{k})-i\gamma)}
-\frac{1}{\omega-(-\omega(\textbf{k})-i\gamma)}\right],
\label{chi2pole}
\end{equation}
where $\omega(\textbf{k})$ is the collective mode frequency and $A=A(\textbf{k})$ is arbitrary. Using Eq. (\ref{chi-tot}) one obtains the proper density response function
\begin{equation}
\overline{\chi}(\textbf{k},\omega)=\frac{A}{2F}\left[\frac{1}{\omega-(\Omega(\textbf{k})-i\gamma)}
-\frac{1}{\omega-(-\Omega(\textbf{k})-i\gamma)}\right],
\label{}
\end{equation}
where
\begin{equation}
\Omega(\textbf{k})=F(\textbf{k})\omega(\textbf{k}),
\label{Omega}
\end{equation}
\begin{equation}
F(\textbf{k})=\left(1-\frac{\varphi(\textbf{k})A(\textbf{k})}{\omega(\textbf{k})}\right)^{\frac{1}{2}}.
\label{}
\end{equation}
$\chi(\textbf{k},\omega)$, being a plus-function, has poles in the lower half of the $\omega$-plane only. In contrast,  $\overline{\chi}(\textbf{k},\omega)$ has now a pair of complex poles
\begin{eqnarray}
z_{1}&=&\Omega(\textbf{k})-i\gamma, \\
z_{2}&=&-\Omega(\textbf{k})-i\gamma
\label{}
\end{eqnarray}
not constrained to the lower half-plane. These poles become pure imaginary when $(1-\varphi A/\omega(\textbf{k}))$ turns negative:
\begin{equation}
z_{1}=i(y-\gamma)
\label{z1}
\end{equation}
\begin{equation}
z_{2}=i(-y-\gamma)
\label{z2}
\end{equation}
with
\begin{equation}
y(\textbf{k})=\omega(\textbf{k})f(\textbf{k}),
\label{y}
\end{equation}
\begin{equation}
f(\textbf{k})=\left(\frac{\varphi(\textbf{k})A(\textbf{k})}{\omega(\textbf{k})}-1\right)^{\frac{1}{2}}.
\end{equation}
$z_2$ is always in the lower half-plane, but $z_1$ may be in the upper half-plane, if the condition
\begin{equation}
\omega(\textbf{k})f(\textbf{k})-\gamma(\textbf{k})>0
\label{vio-cond}
\end{equation}
is satisfied. This we will assume to be the case in the sequel and we will take $\gamma\rightarrow 0$ accordingly.\par

In order to proceed we have to determine the value of the coefficient $A(\textbf{k})$ in Eq. (\ref{chi2pole}). We do this by requiring that Eq. (\ref{chi2pole}) satisfies the static FDT, Eq. (\ref{static-FDT}), which then provides
\begin{equation}
A(\textbf{k})=\beta n_0 S(\textbf{k})\omega(\textbf{k}).
\end{equation}
Then
\begin{align}
F^2(\textbf{k})=-f^2(\textbf{k}) \nonumber
&=1-\beta n_0\varphi(\textbf{k})S(\textbf{k})\\ \nonumber
&=1+\varphi(\textbf{k})\chi(\textbf{k})\\
&=1/\varepsilon(\textbf{k})
\end{align}
This relationship lends itself to the interpretation of $\Omega(\textbf{k})$
as the pressure contribution to the viscoelastic plasma frequency \cite{Vignale}. As the sign of $\varepsilon(\textbf{k})$ is identical to that of the compressibility, it is now obvious that this viscoelastic frequency $\Omega(\textbf{k})$ morphs into the $y-pole$ as the compressibility turns negative. Even though this statement is derived only from the simplified model calculation, there seems to be no doubt that the physical picture it provides is of general validity.\\ \par
A further simplification can be achieved by observing that
the $k\rightarrow 0$ limit of Eq.(\ref{chi2pole}) is equivalent to
\begin{align}
&\chi''(\textbf{k},\omega) \nonumber \\
&=\frac{\pi}{2}\beta n_0 S(\textbf{k})\omega(\textbf{k})[\delta(\omega+\omega(\textbf{k}))-\delta(\omega-\omega(\textbf{k}))],
\end{align}
which may be seen to be tantamount to the classical version of the familiar Feynman Ansatz \cite{Feynman,Feynman-GK} for $S(\textbf{k},\omega)$,
\begin{equation}
S(\textbf{k},\omega)=\frac{1}{2}S(\textbf{k})[\delta(\omega-\omega(\textbf{k}))+\delta(\omega+\omega(\textbf{k}))].
\label{Feynman}
\end{equation}
with its corollary
\begin{equation}
\omega(\textbf{k})=\omega_{\rm p}\frac{1}{\sqrt{S(\textbf{k})}}\frac
{\overline{k}}{\sqrt{3\Gamma}},
\label{model-disp}
\end{equation}
where $\overline{k}=ka$. The values of the $\Omega(\textbf{k})$ pole and of the linked $y(\textbf{k})$ pole are determined from Eqs. (\ref{Omega}) and (\ref{y}):
\begin{equation}
\Omega(\textbf{k})=\omega_{\rm p}\frac{1}{\sqrt{-\overline{\chi}(\textbf{k})/\beta n_0}}\frac
{\overline{k}}{\sqrt{3\Gamma }}
\label{Omega-model}
\end{equation}
\begin{equation}
y(\textbf{k})=\omega_{\rm p}\frac{1}{\sqrt{\overline{\chi}(\textbf{k})/\beta n_0}}\frac
{\overline{k}}{\sqrt{3\Gamma }}
\label{y-model}
\end{equation}
with
\begin{equation}
\overline{\chi}(\textbf{k})=\beta n_0\frac{S(\textbf{k})}{\beta n_0 \varphi(\textbf{k})S(\textbf{k})-1}.
\label{static-chitot}
\end{equation}

At this point we may adopt the language of Section II and split $\overline{\chi}(\textbf{k},\omega)$ into its regular $\Xi(\textbf{k},\omega)$ and anomalous $\Upsilon(\textbf{k},\omega)$ parts. Then we observe that for $\Gamma<\Gamma_{\ast}$ or for $k>k_{\ast}$
\begin{align}
\Xi(\textbf{k},\omega)&=A\omega(\textbf{k})\frac{1}{\omega^2-\Omega^{2}(\textbf{k})} \nonumber \\
&=\omega_{\rm p}^2\beta n_0\frac
{\overline{k}^2}{3\Gamma }\frac{1}{\omega^2-\Omega^{2}(\textbf{k})},\\
\Upsilon(\textbf{k},\omega)&=0,
\end{align}
and for $\Gamma>\Gamma_{\ast}$, $k<k_{\ast}$
\begin{align}
\Xi(\textbf{k},\omega)&=0,\\
\Upsilon(\textbf{k},\omega)&=A\omega(\textbf{k})\frac{1}{\omega^2+y^2(\textbf{k})} \nonumber \\
&=\omega_{\rm p}^2\beta n_0\frac
{\overline{k}^2}{3\Gamma }\frac{1}{\omega^2+y^2(\textbf{k})}.
\label{Upsilon-model}
\end{align}
Comparing Eq. (\ref{Upsilon-model}) with the earlier definition (Eq. (\ref{Upsilon})) of $\Upsilon(\textbf{k},\omega)$ we can express the parameter $Q(\textbf{k})$ as
\begin{align}
Q(\textbf{k})&=\frac{A(\textbf{k})}{f(\textbf{k})} \nonumber \\
&=\beta n_0\omega_{\rm p}\sqrt{\overline{\chi}(\textbf{k})/\beta n_0}\frac{\overline{k}}{\sqrt{3\Gamma}}.
\label{Q-model}
\end{align}

\begin{figure}[htb]
\includegraphics[width=1.0\columnwidth]{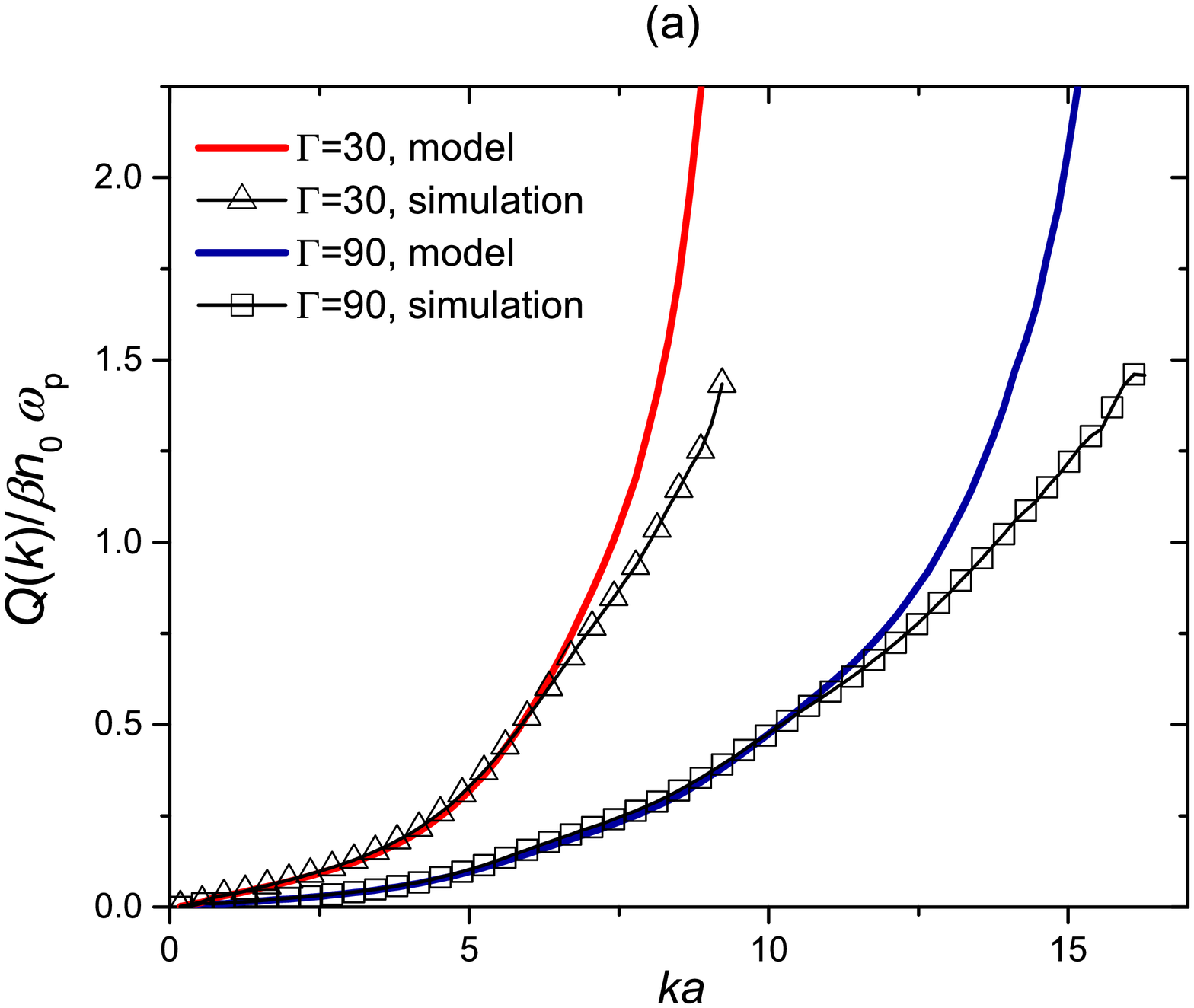}
\includegraphics[width=1.0\columnwidth]{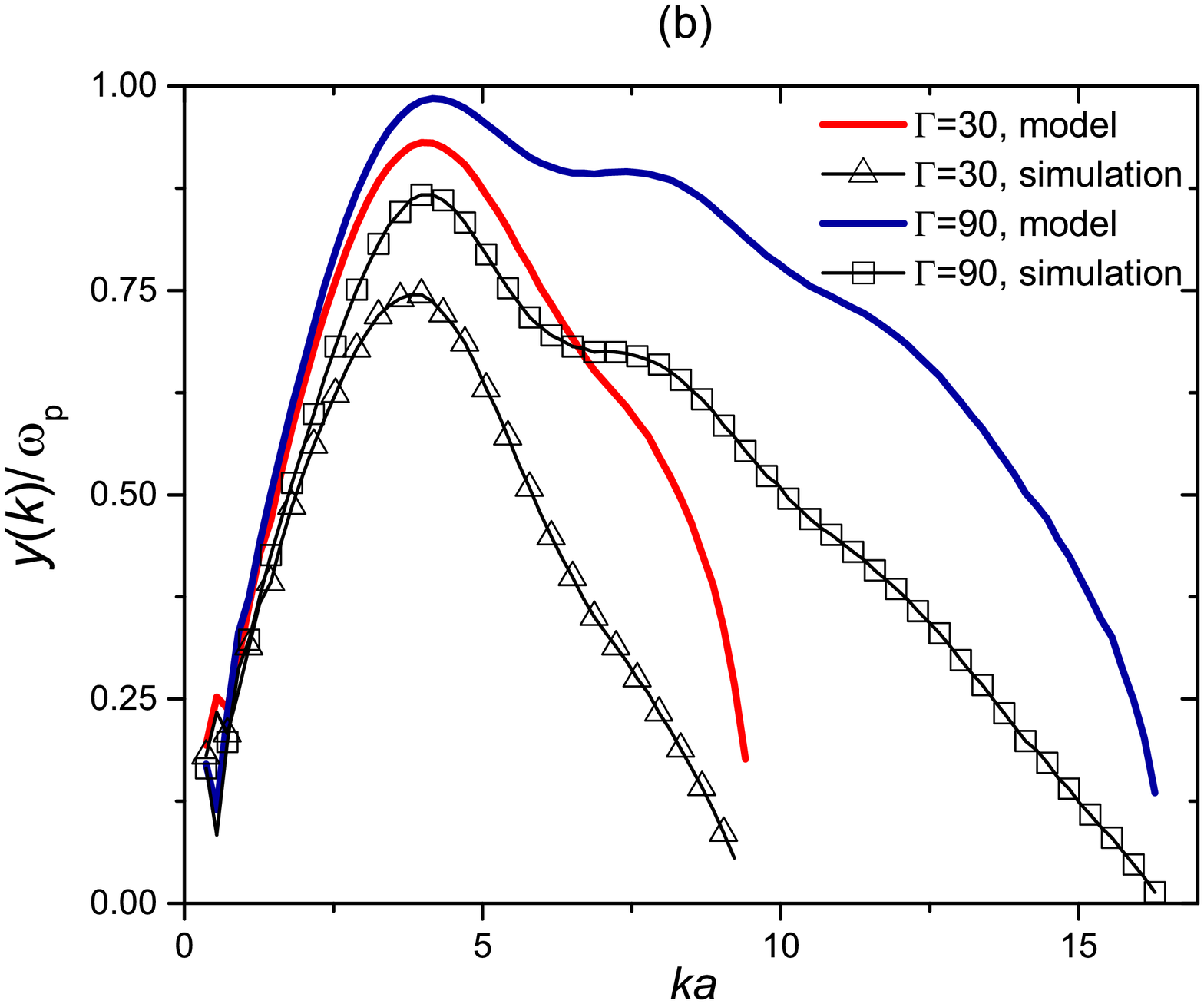}
\caption{Comparison of the $Q(\textbf{k})$ and $y(\textbf{k})$ parameters of the imaginary pole resulting from the model calculation with the exact findings from the MD simulations at $\Gamma$ values indicated.}
\label{Qy-sim-model}
\end{figure}

With the aid of the small-$k$ expansion of $S(\bf k)$
\begin{equation}
S(k\to 0)=\frac{\overline{k}^{2}}{3\Gamma}\left(1-L\frac{\overline{k}^{2}}{3\Gamma}\right),
\label{Sk-exp}
\end{equation}
one can work out, as a by-product, the small-$k$ expansions of $Q(\bf k)$ and $y(\bf k)$ in the expectation that they will serve as a guide for the small-$k$ expansion of the exact expressions.
\begin{equation}
Q(k\to 0)=-\frac{\omega_{\rm p}\beta n_0}{L}(c\overline{k}+\frac{1}{2}c^{3}\overline{k}^{3}),
\label{Q-exp}
\end{equation}
\begin{equation}
y(k\to 0)=\omega_{\rm p}(c\overline{k}-\frac{1}{2}c^{3}\overline{k}^{3}),
\label{y-exp}
\end{equation}
where
\begin{equation}
c=\left(-\frac{L}{3\Gamma}\right)^{\frac{1}{2}}.
\label{c-exp}
\end{equation}
Using the MD data for $S(\textbf{k})$, in Fig.~\ref{Qy-sim-model}, we compare the approximate results (\ref{y-model}) and (\ref{Q-model}) with the exact findings from the MD simulations in Fig.~\ref{vio-parameters}: we see a reasonable agreement, showing that the somewhat unexpected behavior of these quantities is well reproduced by the model calculation. This may be taken as an indication that the physical mechanism identified in the model calculation operates in the exact formalism as well.\\ \par

The essence of the collective behavior discussed in Section III can be easily illuminated by employing the current model. Now, in view of Eqs. (\ref{tr-eps}) and (\ref{Upsilon-model}) $\widetilde{\varepsilon}(\textbf{k},\omega)\equiv  \varepsilon(\textbf{k},\omega)$ and the exact dispersion relation becomes
\begin{equation}
1-\frac{\omega_{\rm p}^2}{\omega^2+y^2(\textbf{k})}=0,
\end{equation}
which doesn't provide any new information beyond Eqs. (\ref{z1})-(\ref{z2}). As to the imaginary part $\overline{\chi}''(\textbf{k},\omega)$, we let $\gamma$ assume a small, but finite value, yielding
\begin{equation}
\overline{\chi}''(\textbf{k},\omega)=-2\beta n_0\omega_{\rm p}^2\gamma(\textbf{k})
\frac{\omega}{(\omega^2+y^2(\textbf{k}))^2}\frac{\overline{k}^2}{3\Gamma}.
\end{equation}
The maximum of this expression is at
\begin{equation}
p(\textbf{k})=y(\textbf{k})/\sqrt{3},
\end{equation}
qualitatively explaining the origin of the structure of Fig.~\ref{peak}.

To summarize the moral of the model calculation, it has elucidated how
the connection between the violation of causality and the negative value of the compressibility emerges. As long as the compressibility of the system stays positive, the pressure generated $\Omega(\textbf{k})$ pole in $\overline{\chi}(\textbf{k},\omega)$ leads to a positive addition to the plasma frequency; once, however, the compressibility turns negative, the pole migrates into the upper half-plane: the existence of such a pole entails the development of the anomalous part of the response, which is the pivotal quantity of the acausal behavior.

\section{TIME DOMAIN}

\begin{figure}[htb]
\includegraphics[width=1.0\columnwidth]{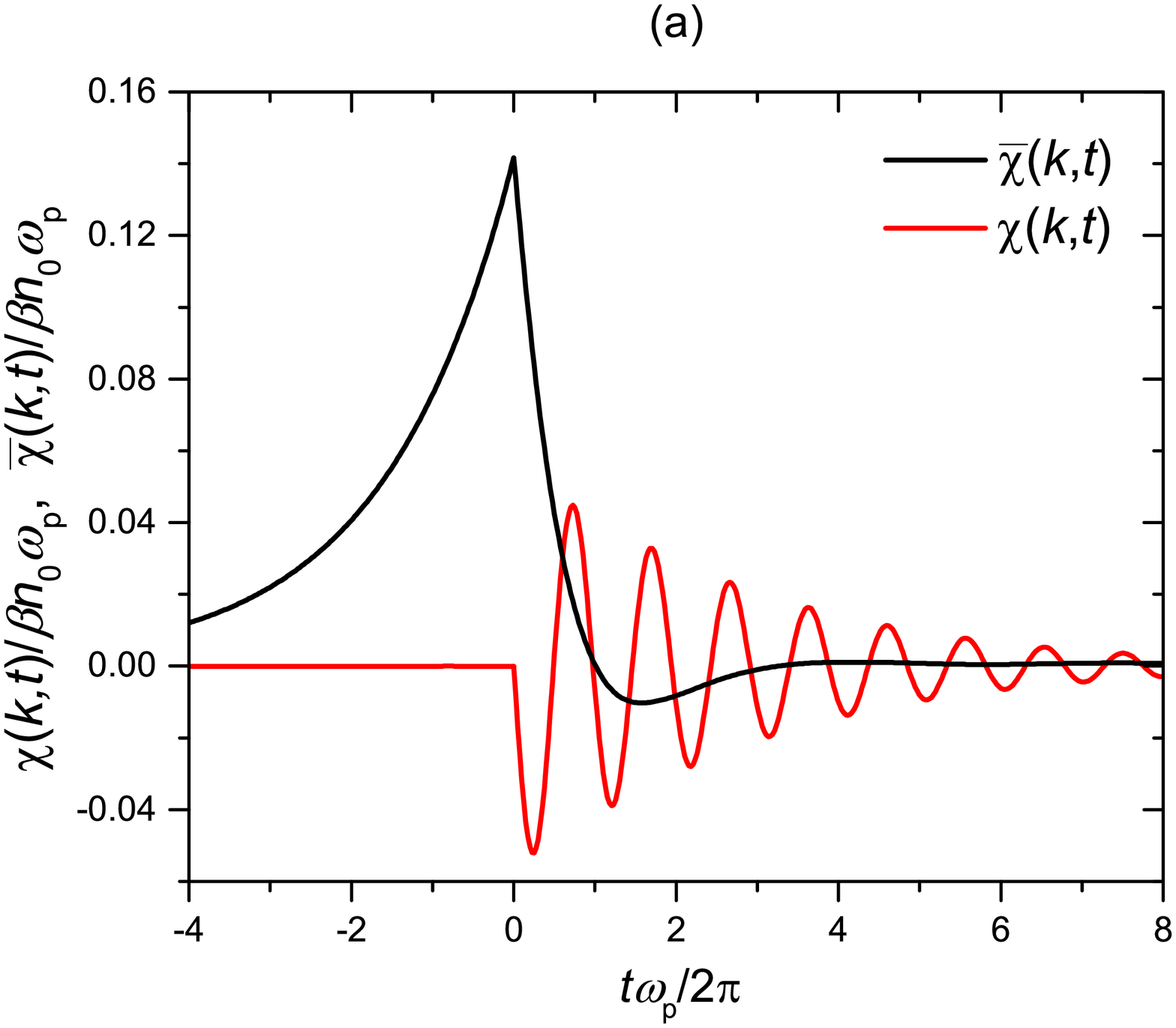}
\includegraphics[width=1.0\columnwidth]{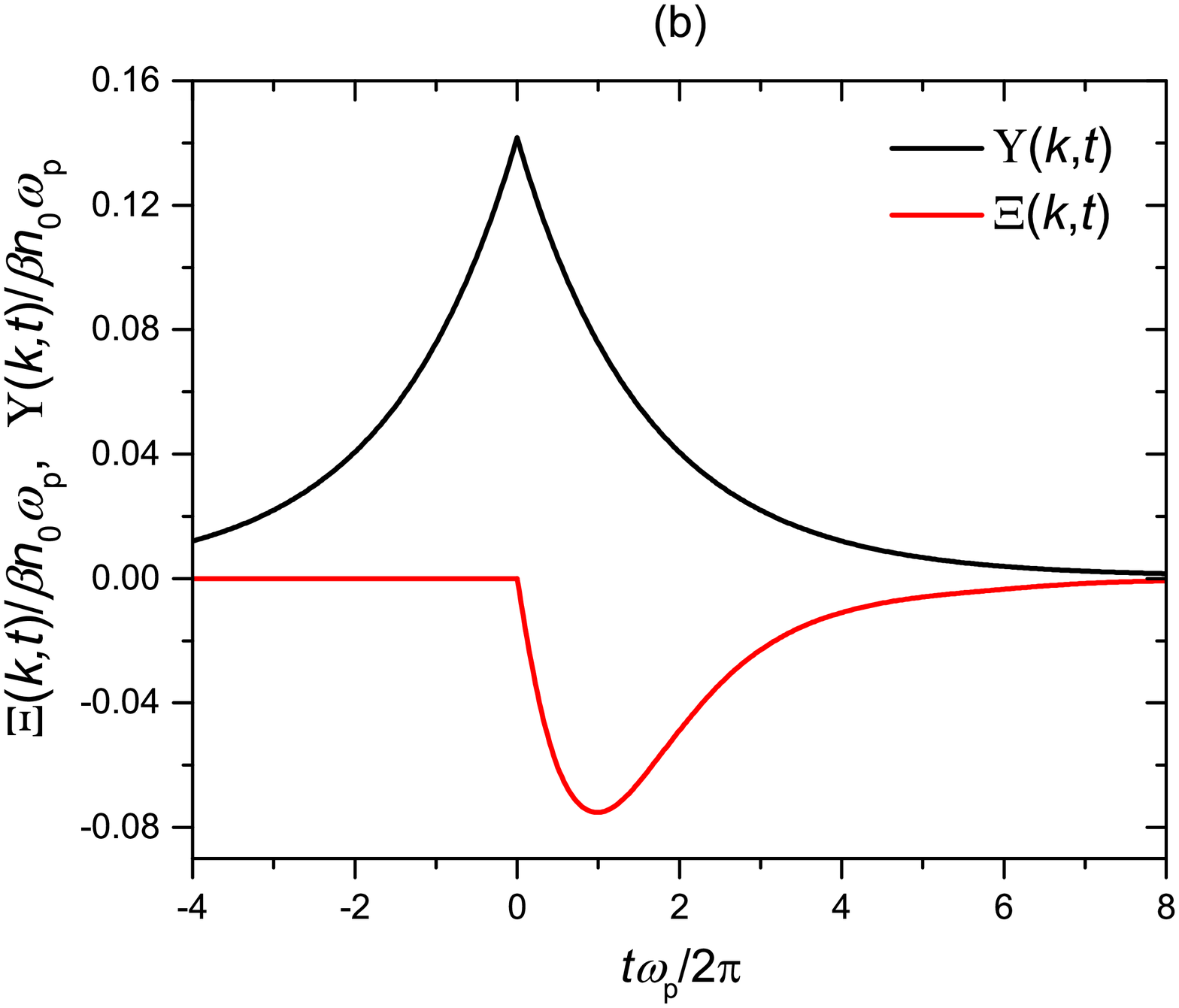}
\caption{(a) Time-dependence of the external and the proper density response functions, $\chi(\textbf{k},t)$ and $\overline{\chi}(\textbf{k},t)$, at $\Gamma=5$ and $ka=0.91$, $k_{\ast}(\Gamma=5)a=3.71$. (b) Split of $\overline{\chi}(\textbf{k},t)$ into the anomalous $\Upsilon(\textbf{k},t)$ and the regular $\Xi(\textbf{k},t)$ terms at the same parameter values .}
\label{time}
\end{figure}

\begin{figure}[htb]
\includegraphics[width=1.0\columnwidth]{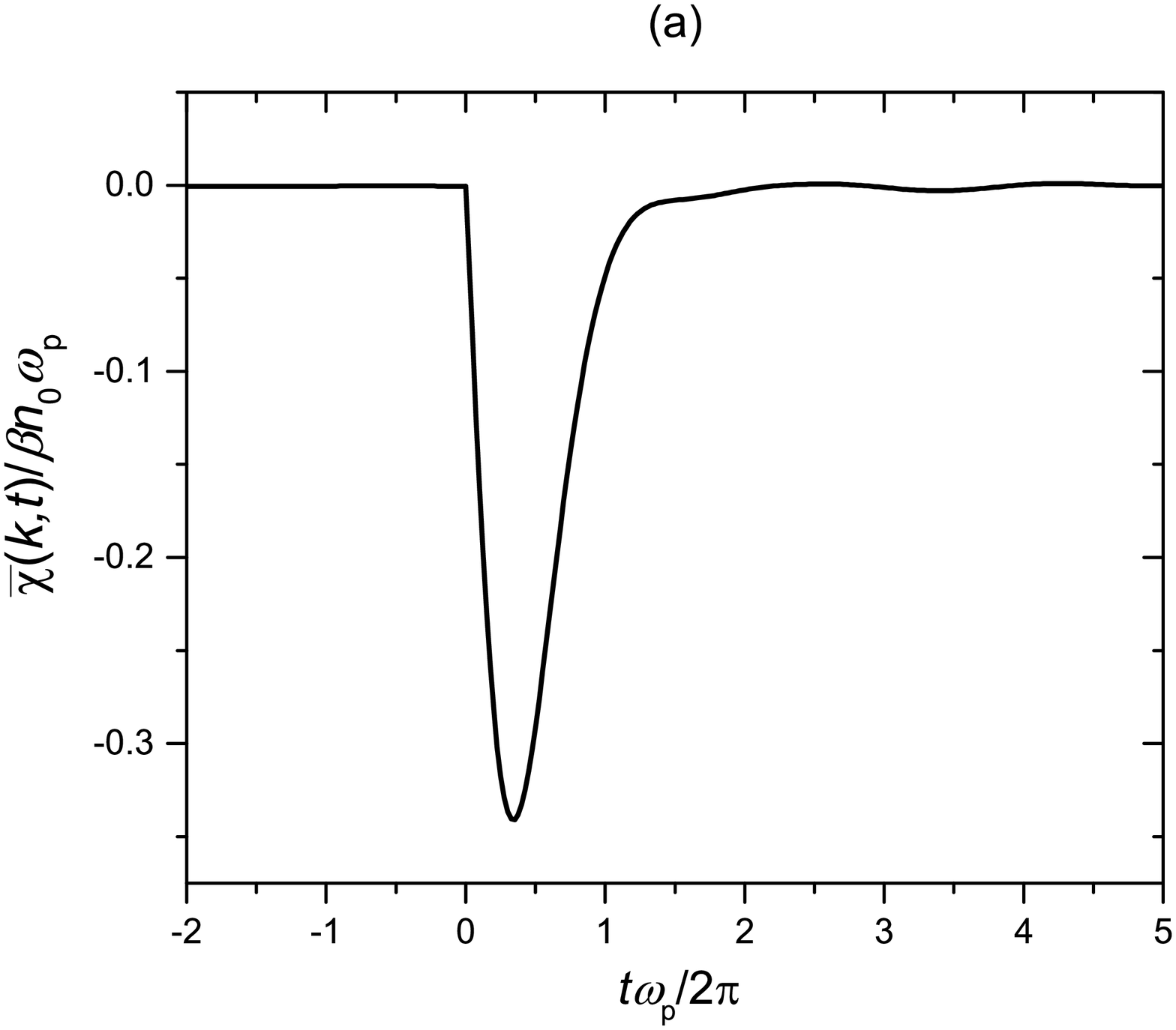}
\includegraphics[width=1.0\columnwidth]{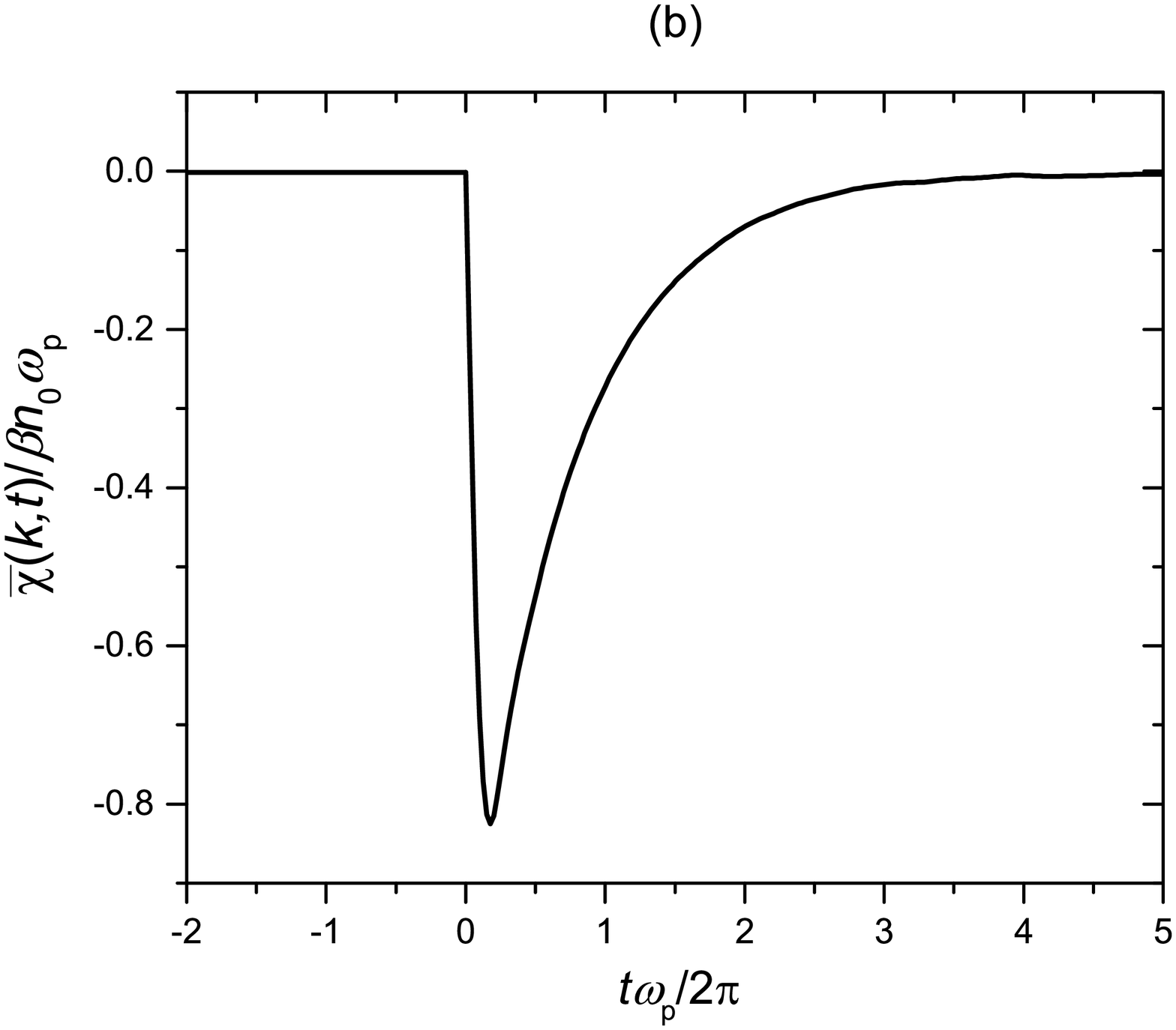}
\caption{Time-dependence of the proper density response function $\overline{\chi}(\textbf{k},t)$ outside of the domain of the violation: (a) at $\Gamma=1<\Gamma_{\ast}$ and $ka=0.91$, (b) at $\Gamma=5>\Gamma_{\ast}$ and  $ka=4.53>k_{\ast}(\Gamma=5)a=3.71 $}
\label{time-novio}
\end{figure}

A deeper insight about the physical processes implied by the acausal behavior can be gained by examining the time domain developments of the various response functions and of  the  related physical quantities. In Fig.~\ref{time}(a) we display the $\chi(\textbf{k},t)$ and $\overline{\chi}(\textbf{k},t)$ time functions, obtained by calculating the inverse Fourier transforms of $\chi(\textbf{k},\omega)$ and $\overline{\chi}(\textbf{k},\omega)$.
It is evident  that they possess the expected time behavior: the external $\chi(\textbf{k},t)$ is causal, i.e. it vanishes for all $t<0$ values, while the proper
$\overline{\chi}(\textbf{k},t)$ is acausal, carrying an exponential tail in the $t<0$ domain. This anomalous part is the offspring of the $y$-pole, discussed above. To see this clearly, we have separated in Fig.~\ref{time}(b) the anomalous part from the well-behaved regular, causal contribution. Fig.~\ref{time-novio}(a) and Fig.~\ref{time-novio}(b) also show that once we are outside the negative compressibility domain, either by reducing $\Gamma$ below its $\Gamma_{\ast}$ value, or by increasing $k$ over its $k_{\ast}$ value, $\overline{\chi}(\textbf{k},t)$ recovers its normal causal behavior.

\begin{figure*}[htb]
\centering
\includegraphics[width=20cm]{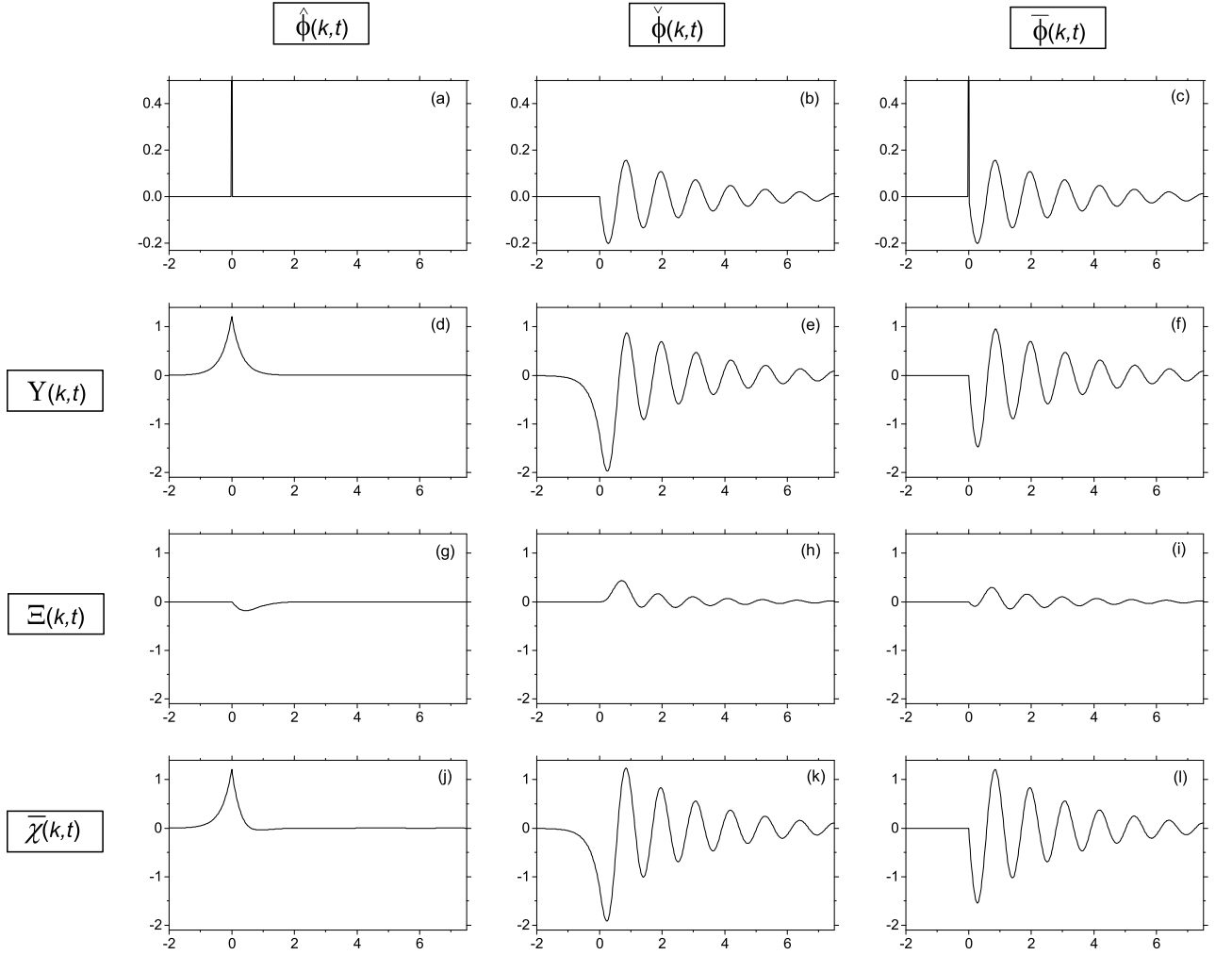}
\caption{Details
of the time evolutions of various responses generated by the proper density response function in the violation domain.  The columns of the graphic matrix represent  the external $\widehat{\Phi}$, polarization $\widecheck{\Phi}$ and total $\overline{\Phi}$ perturbing fields. The entries in the top row in panels (a), (b) and (c) illustrate the column headings. The rows are labelled by $\bar{\chi}(\textbf{k},t)$ and its regular $\Xi(\textbf{k},t)$ and anomalous $\Upsilon(\textbf{k},t)$ components. An entry in the matrix, in panels (d) though (l) displays the convolution product of the respective   density responses and perturbing fields. The physically measurable density  perturbation $n(\textbf{k},t)$ is the last entry, panel (l), with an obviously causal behavior. All the entries with non-vanishing contributions for $t<0$ represent auxiliary entities, without any physical observability. $\Gamma=90$, $ka=1.81$.}
\label{conv-g90}
\end{figure*}

\begin{figure*}[htb]
\centering
\includegraphics[width=20cm]{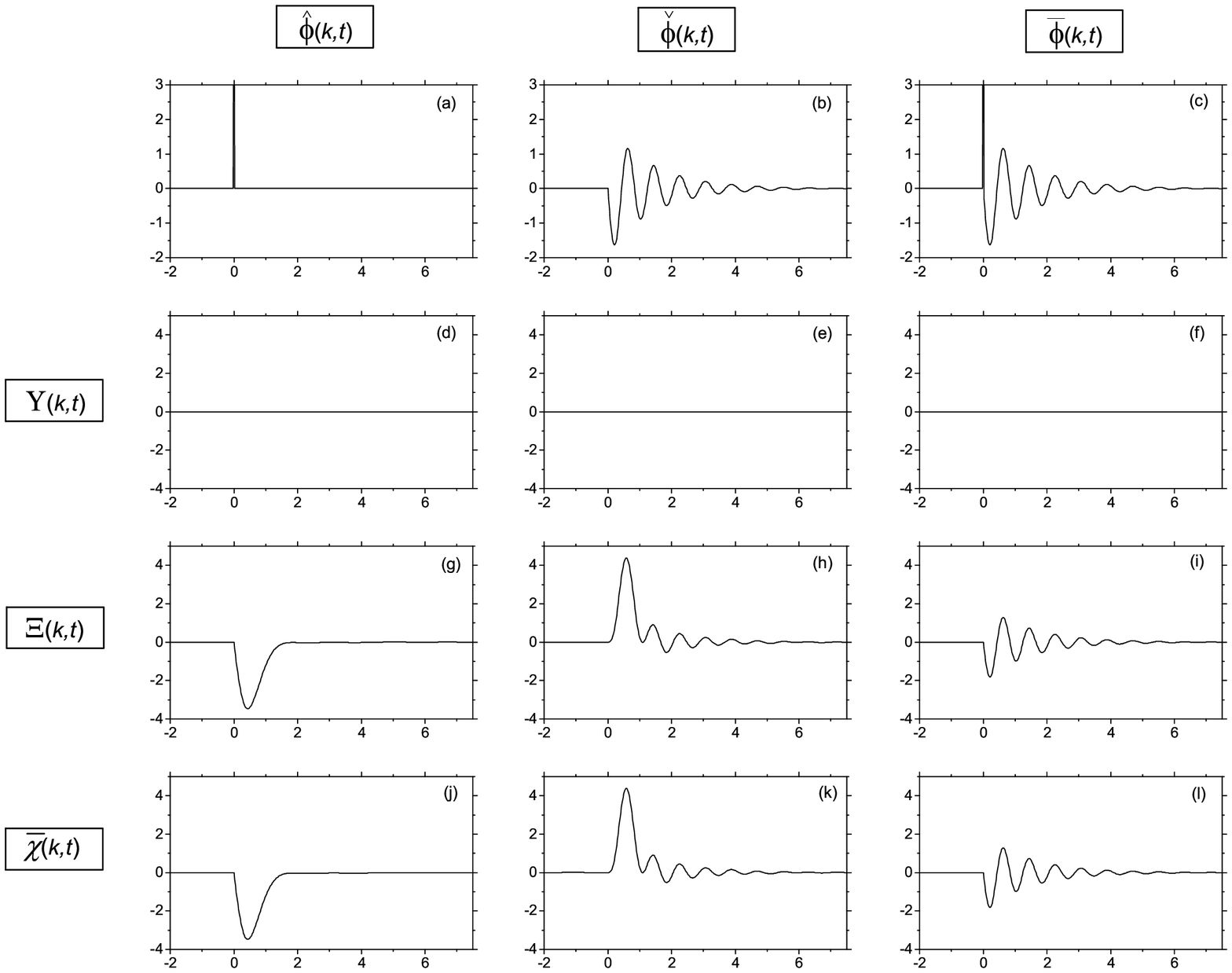}
\caption{Same as Fig. 14 outside of the violation domain. $\Gamma=1$, $ka=0.72$.}
\label{conv-g1}
\end{figure*}

There are two measurable physical quantities in the current scenario: the total perturbed (induced) density $n(\textbf{k},t)$ and the total (external plus induced) potential $\overline{\Phi}(\textbf{k},t)=\widehat{\Phi}(\textbf{k},t)+\widecheck{\Phi}(\textbf{k},t)$. These are necessarily causal quantities. Nevertheless, the connection between them is expressed through the acausal $\overline{\chi}(\textbf{k},t)$
\begin{equation}
n(\textbf{k},t)=\overline{\chi}(\textbf{k},t)\ast\overline{\Phi}(\textbf{k},t).
\end{equation}
(The $\ast$ sign designates a convolution integral). On the other hand, neither of the two (unphysical) constituents $\overline{\chi}(\textbf{k},t)\ast\widehat{\Phi}(\textbf{k},t)$ and
$\overline{\chi}(\textbf{k},t)\ast\widecheck{\Phi}(\textbf{k},t)$ are causal functions. The intricate relationships that render the combination of the various -not obviously, but in fact unphysical- quantities causal is illustrated in Fig.~\ref{conv-g90}. The figure follows the simple sequence ensuing from a
$\delta$-function perturbation by an external potential
\begin{equation}
\widehat{\Phi}(\textbf{k},t)=\Phi_0(\textbf{k})\delta(t),
\end{equation}
as it generates induced potential and induced density responses
both via the regular (causal) and
anomalous (acausal) contributions.
It is instructive to compare their behavior
with the one shown in the companion Fig.~\ref{conv-g1} for the domain of the normal (causal) response. The most flagrant difference may be observed in panels (j) and (k), portraying the separate "system responses" to the external and polarization fields. One would be inclined to believe that they are indeed separable and measurable quantities. A little reflection shows though that this is not the case. Thus their  strange acausal behavior one is confronted with in these figures shouldn't be, in fact, of any physical consequence. Through these figures one may also be able to glean the details of what has been pointed out above as to the linkage of two causal quantities via an acausal response.

\section{SHORT RANGE VS. LONG RANGE}

As it has been clear from the outset, it is the negative compressibility that is ultimately responsible for the acausal behavior. The system we have focused on, the Coulombic OCP does exhibit a negative compressibility. But the OCP
is not a self-consistent model: its thermodynamic stability is maintained only by the rigid compensating background. In fact, all physically realizable systems must have positive compressibility  to prevent them from collapse. Thus one may wonder whether the acausality phenomenon would survive in a more realistically described system, in which the long range Coulomb interaction is replaced by a short range potential that does not require the support of a rigid background. An immediate example is the Yukawa OCP that has been widely used in the recent literature in the context of complex plasmas, colloids and white dwarf interiors (see, e. g. \cite{Y-example1,Y-example2,Y-example3}). Here the  interaction is described by the finite $\mu^{-1}$ range exponential Yukawa potential, whose Fourier representation is
\begin{equation}
\varphi(\textbf{k})=\frac{4\pi Z^{2}e^{2}}{k^2+\mu^{2}}
\end{equation}
We now examine to what extent the previously derived relationships are affected by the change from the long range Coulomb to  the finite range Yukawa potential.
All the fundamentals, Eq. (1) through (\ref{static-FDT}) are interaction independent and still remain valid. However, a careful re-interpretation of Eq. (\ref{compsumrule}) is required. It should be recalled that the entirely model independent derivation of
the compressibility sum rule rests upon the static FDT, Eq. (\ref{static-FDT})
and the thermodynamic relationship between $S(\textbf{k})$ and the compressibility for finite range (but not Coulomb) interactions \cite{Hansen-book}:
\begin{equation}
S(k=0)=1/L.
\end{equation}
Here $L$ is the total physical inverse compressibility, which  includes, in addition to the correlational contribution, a Hartree-term $L_{\rm H}=3\Gamma/\bar{\mu}^2$, $\bar{\mu}=\mu a$. The correlational $L_{\rm corr}$ is always negative, while $L_{\rm H}$ is positive, ensuring that the total inverse compressibility,
\begin{equation}
L=1+L_{\rm corr}+L_{\rm H},
\end{equation}
always stays positive. It is now this compressibility that governs the small-$k$ behavior of $\chi(\textbf{k})$:
\begin{equation}
\chi(k=0)=-\beta n_0/L
\end{equation}
Proceeding now to the calculation of $\overline{\chi}(\textbf{k},\omega)$ following Eq. (\ref{chi-tot}), yields the entirely different static behavior:
\begin{equation}
\overline{\chi}(k=0)=-\beta n_0/\mathcal{L},
\end{equation}
where $\mathcal{L}=1+L_{\rm corr}$. This remarkable  result is the consequence of the cancellation of the Hartree compressibility by the $k=0$ value of the interaction potential.

Also, calculating $\varepsilon(\textbf{k})$ we find
\begin{equation}
\varepsilon(k=0)=1+\frac{3\Gamma}{\mathcal{L}}\frac{1}{\bar{\mu}^{2}}=\frac{L}{\mathcal{L}}.
\end{equation}
Now it is clear that $\chi(\textbf{k},\omega)$
even though in its small-$k$ behavior differs substantially
from its Coulombic counterpart, will exhibit a normal causal behavior, as it must; on the other hand, $\overline{\chi}(\textbf{k},\omega)$ quite in line with the Coulombic case, will be acausal, once $\mathcal{L}<0$, which does happen, similarly to the OCP, for some $\Gamma_{\ast}(\mu)> \Gamma_{\ast}$ \cite{Khrapak-1,Khrapak-2}.\par

Thus, whatever has been said about the acausal behavior of
$\overline{\chi}(\textbf{k},\omega)$ for the Coulombic OCP so far
stands qualitatively
for the Yukawa OCP as well, short range character of the interaction notwithstanding. Quantitatively of course, in the absence of relevant calculations of the response functions not much can be stated. Some inferences, though, can be drawn from the two-pole model, presented in Section IV, after the correct re-interpretation of $\varphi(\textbf{k})$ and $\omega(\textbf{k})$. One may represent this latter through the Feynman relation Eq. (\ref{Feynman}) as well, leaving Eqs. (\ref{model-disp}) through (\ref{static-chitot}) formally invariant. Nevertheless, when
$\overline{\chi}(\textbf{k})$ is calculated as an explicit function of $\textbf{k}$
the resulting expression turns out to be quite complex, which we do not find useful to display here, restricting ourselves to the analysis of $k\to 0$ limit only.

The long wavelength collective excitation in the Yukawa OCP is an acoustic plasmon mode entangled with the hydrodynamic sound \cite{Luciano}:
\begin{align}
\omega(\textbf{k})&=\omega_{\rm p}\sqrt{\frac{L}{3\Gamma}}\overline{k},
\end{align}
while the pole frequency $\Omega(\textbf{k})$ remains similar to (\ref{Omega-model}):
\begin{align}
\Omega(\textbf{k})&=\omega_{\rm p}\sqrt{\frac{\mathcal{L}}{3\Gamma}}\overline{k}.
\end{align}
Using these results and proceeding now to the determination of $Q$ and $y$ according to (\ref{Q-model}) and (\ref{y-model}) yields formulas, which except for the replacement of $L$ by $\mathcal{L}$, are identical to Eqs. (\ref{Q-exp}) through (\ref{c-exp}). This similarity exhibited by the $y$-poles
of the Coulomb and Yukawa OCP-s, their differences notwithstanding, is quite noteworthy.

In conclusion, we note that similarly to the Coulombic OCP, high quality MD simulation data on the equilibrium fluctuation spectra (dynamical structure function, etc.) are available for the Yukawa OCP as well \cite{Arkhipov,Mithen2,Hanno,Luciano}. What is missing at the present time is the conversion of these data into a formulation of the response functions along the pattern set by this paper. This has to await future work.

\section{CONCLUSIONS}

The improvements in computer Molecular Dynamics  simulations of the equilibrium dynamics of many-body systems over the past decade   has resulted in the accumulation of high quality data on the dynamical structure function  $S(\textbf{k},\omega)$ for  Coulomb-like systems, in particular for the one-component plasma (OCP) over a broad range of $\Gamma$ coupling values. These data constitute a depository of a wealth information on the system, on  the real and imaginary  parts of a family of linear response functions  in particular.  With the aid of the Fluctuation-Dissipation Theorem (FDT) and the Kramers-Kronig  (KK) relations the data  can be converted into a catalogue of the detailed behavior of these functions. In this paper, we have used results of this approach to analyze  the density response functions and the dielectric response function of the OCP  in the domain of strong coupling, which has hitherto been inaccessible either to analytic or to direct computational methods. Our main focus is directed to the phenomenon that has attracted a great deal of interest and has created some controversy for some time, namely the apparent acausal behavior of some of the response functions, manifested by the violation of the KK relations in this domain.
It has been clear from the outset that the negative value of the static response function, due to the compressibility becoming negative for $\Gamma>3$, triggers the acausal behavior.

The game changing significance  of the onset of negative compresibility (or, equivalently, of the static dielectric function $\varepsilon(\textbf{k})$ assuming negative values) in many respects  was emphasized by Krizhnits, Dolgov and collaborators in a series of publications \cite{Kirzhnits67,Kirzhnits81,Dolgov-Maksimov,Dolgov-supercond} since the 1960-s. Nevertheless, none of these authors, -or, the best of our knowledge, no other research groups either (see, however, the very recent work  \cite{Vorberger2})- have explored the consequences  of this feature on the dynamical properties of the response functions. This, of course, to a great extent has been due to the lack of available data on which such an analysis could have been based. It is  now in this work that we have been able to create a full picture of the evolution of the response functions in the anomalous domain.

Having determined the  maximum $k_{\ast}(\Gamma)$, forming the boundary of the acausal region  we have given a detailed picture of the response function, as it  splits into a "regular" (KK-preserving) and an "anomalous" (KK-violating) part, the latter being generated by the so-called $y-pole$  in the upper $\omega$ half-plane. The existence of such a pole as a concomitant to the acausal behavior is well established  on mathematical grounds in the literature; here we have been able to trace its physical origin to the migration of  the known viscoelastic (hydrodynamic) pole as the compressibility assumes a negative value.

The knowledge of the dielectric response function, $\varepsilon(\textbf{k},\omega)$, has made it possible to analyze the plasmon dispersion in the acausal domain in a less ambiguous manner than via the usual method of examination of the peaks of the dynamical structure function $S(\textbf{k},\omega)$. (For a recent discussion on the issues involved see \cite{Vorberger}.) The dispersion relation so displayed clearly  exhibits a pronounced weakly damped roton minimum, which seems to qualify as a \textit {bona fide} collective excitation. We have demonstrated that its existence is linked to the anomalous part of the response function. This observation has led us to the conclusion that the roton minimum is the consequence of the negative compressibility, a statement that we believe to be of quite general system independent validity.

In order to understand better what the acausal behavior actually means in terms of time dependence, we have complemented our study of the response by analyzing its Fourier transform back into the time domain. There are two major conclusions we could draw from this study. Our first remark concerns the observability of the acausal behavior, i.e., the question whether there exists a physical quantity in whose time evolution any activity prior to the perturbation can be detected. The answer, perhaps trivially, is negative. What remains to be understood is how this assertion can be made compatible with
the non-vanishing of the response function for negative times. To see this, one can observe that even though the anomalous part of the response function creates a precursor response to the $\widehat{\Phi}$  perturbation imposed, it does the same for the polarization field $\widecheck{\Phi}$
as well. As a result, the two precursor fields exactly cancel each other, leaving no contribution to the physical response at negative times.
Our second observation concerns the argument, emphasized by Kirzhnits and others, that the lack of causality of the response with respect to the polarization field, in contrast to the external field, is permissible. While this reasoning is certainly correct, it should not be confused with the view that seems to have been implied by some of the discussions on the topic, namely that such an acausal behavior with respect to the polarization field $\widecheck{\Phi}$ is not only permissible but is even physically reasonable. This would happen because $\widecheck{\Phi}$  represents a delayed response with respect to the external perturbation $\widehat{\Phi}$, an activity  of $n$ preceding it, could still be subsequent to $\widehat{\Phi}$. Since, however,  as it can easily be seen, $n$ and $\widecheck{\Phi}$
are simultaneous, this scenario cannot hold. In fact, the precursor time dependences of  the responses are entirely governed by the $y-pole$ and are  not associated with  the time behavior of any physically observable quantity.

The anomalous behavior and the related structural features at strong coupling we have identified in this work are probably characteristic for a wider class of physical systems - classical and quantum - than the OCP studied here. We have shown through the example of the Yukawa OCP that the pivotal role of the  negative compressibility, a unique feature of the OCP, is, in fact, not a restrictive condition: in systems with a physically required positive compressibility it is the negative correlational part of the compressibility that takes over this role.

It should be clearly understood that the manifestations of the anomalous - i.e the dramatically  and sometimes unexpectedly novel - behavior of the response functions once the  critical $\Gamma_{\ast}$ coupling value is exceeded have been established via computer simulations and that there is no cogent theoretical framework behind them. Logical and analytical requirements, such as sum rules, have served as anchors ensuring that the results were built on solid physical foundation, but what they did not provide is a coherent analytic structure from which such results  could have been derived. Starting with the compressibility sum rule as an example,  relying on the behavior of the equilibrium static structure factor it determines the anomalous positive value of the static density response function. But, at the same time, we are missing a consistent analytic expression for the response function that would directly deliver this
result. A more profound demonstration of our lack of knowledge is that the  development of the crucially important $y-pole$ is not a part of any existing response function model. How an analytic formalism appropriate for strong coupling capable of reproducing our results is a profound and important question that evidently we have not tackled in this work. It seems to us very unlikely that the popular local field formalism would be the one to provide the right approach to do this job.

\acknowledgments

ZD thankfully acknowledges funding from the Hungarian National Office for Research, Innovation, and Technology (NKFIH), via grant K134462.

\end{document}